\documentclass[prl,showpacs,twocolumn,superscriptaddress,longbibliography]{revtex4-1}
\usepackage{graphicx}
\usepackage{color}
\usepackage{amssymb}
\usepackage{amsmath}
\usepackage{siunitx}
\usepackage[caption=false]{subfig}
\usepackage{marvosym} % MWM  dingbats for \Laserbeam
\usepackage{natbib} % MWM  natbib

\newcommand{\be}{\begin{equation}}
\newcommand{\ee}{\end{equation}}
\newcommand{\bea}{\begin{eqnarray}}
\newcommand{\eea}{\end{eqnarray}}

\newcommand{\expect}[1]{\left<#1\right>}

\newcommand{\nn}{\nonumber \\ }

\newcommand{\nne}{\nonumber \\ & = &}

\newcommand{\nnp}{\nonumber \\ & & +}

\newcommand{\var}{{\rm var}}

\usepackage{hyperref}

\usepackage{multibib}
\newcommand{\mybibhead}{\vspace{-1cm}}
\newcites{SOM}{\mybibhead}

\newcommand{\vis}{\begin{cal} V \end{cal}}

\definecolor{mygreen}{rgb}{0,0.5,0} 
\definecolor{myred}{rgb}{0.75,0,0} 
\definecolor{myblue}{rgb}{0,0,0.75} 
\definecolor{mymagenta}{cmyk}{0,1,0,0.12} 
\definecolor{mycyan}{cmyk}{1,0,0,0.12} 
\definecolor{myorange}{rgb}{1,0.5,0}

\newcommand{\rtext}[1]{{\color{myred}#1}}

\newcommand{\xor}{\textsc{xor}}

\newcommand{\Pred}{{\cal P}}

\newcommand{\ftime}{\tau_{\rm f}}

\newcommand{\pS}{p_S}
\newcommand{\pL}{p_L}
\newcommand{\vSbar}{\overbar{\vS}}
\newcommand{\vLbar}{\overbar{\vL}}
\renewcommand{\vSbar}{{\langle \vS \rangle}}
\renewcommand{\vLbar}{{\langle \vL \rangle}}
\newcommand{\vcbar}{{\langle \vc \rangle}}

\newcommand{\vref}{v_{\rm ref}}

\newcommand{\vD}{v_{\rm PD}}
\newcommand{\vhang}{v_{\rm h/o}}
\renewcommand{\vhang}{v_{\rm H}}

\newcommand{\vS}{v_{\rm S}}
\newcommand{\vL}{v_{\rm L}}

\newcommand{\vc}{v_{\rm c}}
\newcommand{\vO}{v_{\rm O}}

\newcommand{\vphi}{v_{\phi}}

\newcommand{\vrange}{\Delta v_{\phi}}

\newcommand{\hnow}{h_{\rm f}}
\newcommand{\hprior}{h_{\rm s}}
\newcommand{\tauPD}{\tau_{\rm PD}}

\newcommand{\dik}{D_i^{(k)}}
\renewcommand{\dik}{D_{i,k}}

\newcommand{\corrdist}{k}
\newcommand{\corrvar}{x}

\usepackage{array}
\newcolumntype{L}[1]{>{\raggedright\let\newline\\\arraybackslash\hspace{0pt}}m{#1}}
\newcolumntype{C}[1]{>{\centering\let\newline\\\arraybackslash\hspace{0pt}}m{#1}}
\newcolumntype{R}[1]{>{\raggedleft\let\newline\\\arraybackslash\hspace{0pt}}m{#1}}

\newcommand{\holvar}{\var_{\rm H}}
\newcommand{\prederr}{excess predictability}
\newcommand{\Prederr}{Excess predictability}
\newcommand{\prederrs}{excess predictabilities}

\begin{document}

\title{
%Quantum random numbers with fast randomness extraction for nanoerrors in nanoseconds  \\
%Real-time randomness extraction for nanoerror  random bits on nanosecond budgets \\
%Nanosecond  extraction of  quantum randomness.
%Real-time extraction of quantum randomness: exponential error reduction in nanoseconds\\
{Generation of fresh and pure random numbers for loophole-free Bell tests} \\
%\btext{Real-time extraction of bounded quantum randomness:  nanoerrors in nanoseconds}
%\btext{Real-time extraction of spontaneous emission randomness:  nanoerrors in nanoseconds}
%\rtext{CONFIDENTIAL : DO NOT DISTRIBUTE} \\
%{Generation of fresh and pure random numbers for loophole-free Bell tests}
%\\{\small 90 characters, max, including spaces}
}

\author{Carlos Abellan}
\affiliation{ICFO -- Institut de Ciencies Fotoniques, The Barcelona Institute of Science and Technology, 08860 Castelldefels (Barcelona), Spain}

 \author{Waldimar Amaya}
\affiliation{ICFO -- Institut de Ciencies Fotoniques, The Barcelona Institute of Science and Technology, 08860 Castelldefels (Barcelona), Spain}

\author{Daniel Mitrani}
\affiliation{ICFO -- Institut de Ciencies Fotoniques, The Barcelona Institute of Science and Technology, 08860 Castelldefels (Barcelona), Spain}

\author{Valerio Pruneri}
\affiliation{ICFO -- Institut de Ciencies Fotoniques, The Barcelona Institute of Science and Technology, 08860 Castelldefels (Barcelona), Spain}
\affiliation{ICREA -- Instituci\'{o} Catalana de Recerca i Estudis Avan\c{c}ats, 08015 Barcelona, Spain}

\author{Morgan W. Mitchell}
\affiliation{ICFO -- Institut de Ciencies Fotoniques, The Barcelona Institute of Science and Technology, 08860 Castelldefels (Barcelona), Spain}
\affiliation{ICREA -- Instituci\'{o} Catalana de Recerca i Estudis Avan\c{c}ats, 08015 Barcelona, Spain}

\date{\today}

%\begin{abstract}
%\end{abstract}

\newcommand{\foc}{FoC}
\newcommand{\etal}{{\it et al.}~}

\begin{abstract}

We demonstrate extraction of randomness from spontaneous-emission events less than $\SI{36}{ns}$ in the past, giving output bits with {\prederr} below $10^{-5}$ and strong metrological randomness assurances.  This randomness generation strategy satisfies the stringent requirements for unpredictable basis choices in current ``loophole-free Bell tests'' of local realism 
[Hensen et al., Nature (London) 526, 682 (2015); Giustina et al., Phys. Rev. Lett. 115, 250401 (2015); Shalm et al., Phys. Rev. Lett. 115, 250402 (2015)].
%{[Hensen \etal Nature {\bf 526}, 682 (2015)], Giustina \etal  Phys. Rev. Lett. ms. LL14789 (forthcoming), Shalm \etal Phys. Rev. Lett. ms LL14790 (forthcoming)].
\end{abstract}

\maketitle

Quantum nonlocality \cite{BellP1964} is one of the most striking predictions to emerge from quantum theory.  Beyond their fundamental interest, loophole-free Bell tests enable powerful ``device independent'' information protocols, guaranteed by the impossibility of faster-than-light communication \cite{AcinPRL2007}.  
Bell tests and device-independent protocols employ spacelike separation of measurements to guarantee the nonlocality of correlations \cite{WeihsPRL1998,RoweN2001,ScheidlPNAS2010,GiustinaN2013, ChristensenPRL2013,ErvenNPhot2014b} and the monogamy of correlations under the no-signaling principle \cite{PironioN2010,ColbeckNPhys2012,LoPRL2012}.  To be secure, they must close two space-time loopholes: no basis choice may influence a distant particle (locality loophole), and the entanglement generation must not influence the basis choices (freedom-of-choice~loophole). 
Current efforts \cite{NISTBeacon,HofmannS2012,BernienN2013,GiustinaN2013, ChristensenPRL2013} to simultaneously close the detection \cite{RoweN2001,GiustinaN2013, ChristensenPRL2013}, locality \cite{WeihsPRL1998}, and freedom-of-choice (\foc) \cite{ScheidlPNAS2010,ErvenNPhot2014b} loopholes require random number generators (RNGs) with an unprecedented combination of speed, unpredictability, and confidence \cite{WittmannNJP2012, Brunner:2013br,KoflerARX2014}. 

Here we combine ultrafast RNG by accelerated laser phase diffusion \cite{JofreOE11,AbellanOE2014,Yuan:2014co} with real-time randomness extraction and metrological randomness assurances \cite{MitchellPRA2015} to produce a RNGs suitable for loophole-free Bell tests.  {Because the laser phase diffusion is driven by effects, including spontaneous emission, that are unpredictable both in quantum theory and in an important class of stochastic hidden variable theories, the source can be used to address the ``freedom-of-choice'' loophole \cite{KoflerARX2014b}.  Using a detailed and validated model of the signal generation process, we show the effectiveness of parity-bit randomness extraction of this source. } Under paranoid assumptions, we infer {\prederr}~below ${10^{-5}}$ at $6 \sigma$ statistical confidence for output based on phase-diffusion events less than ${36}$ ns old.  {A statistical analysis based on 2.3 Tbits of random data supports} the metrological assessment of extreme unpredictability.  The results enable definitive nonlocality experiments and secure communications without the need for trusted devices \cite{PironioN2010,LoPRL2012,LiuPRL2013,MizutaniSR2014}. 

As shown in Fig.~\ref{Fig:SpaceTimeDiagram}, the locality and freedom-of-choice loopholes can be closed by spacelike separation of the random events that determine the basis choice {from the distant detection and from the production of the pairs of particles,} respectively \cite{ColbeckNPhys2012}. This {requires generation of randomness in a time window shorter than the light time between the detectors.} Closing the ``detection loophole'' requires high efficiency and motivates protocols very sensitive to predictability of the basis choices.  {Both experiments employing 100\% efficient ``event-ready'' detection \cite{HensenN2015} and those employing high-efficiency photodetection \cite{GiustinaPRL2015, ShalmPRL2015}, are expected to require  {\prederrs} $\epsilon$ below a few times $10^{-5}$ \cite{KoflerARX2014b}. } 

\begin{figure}[!bp]
\centering
\includegraphics[width=0.7\linewidth]{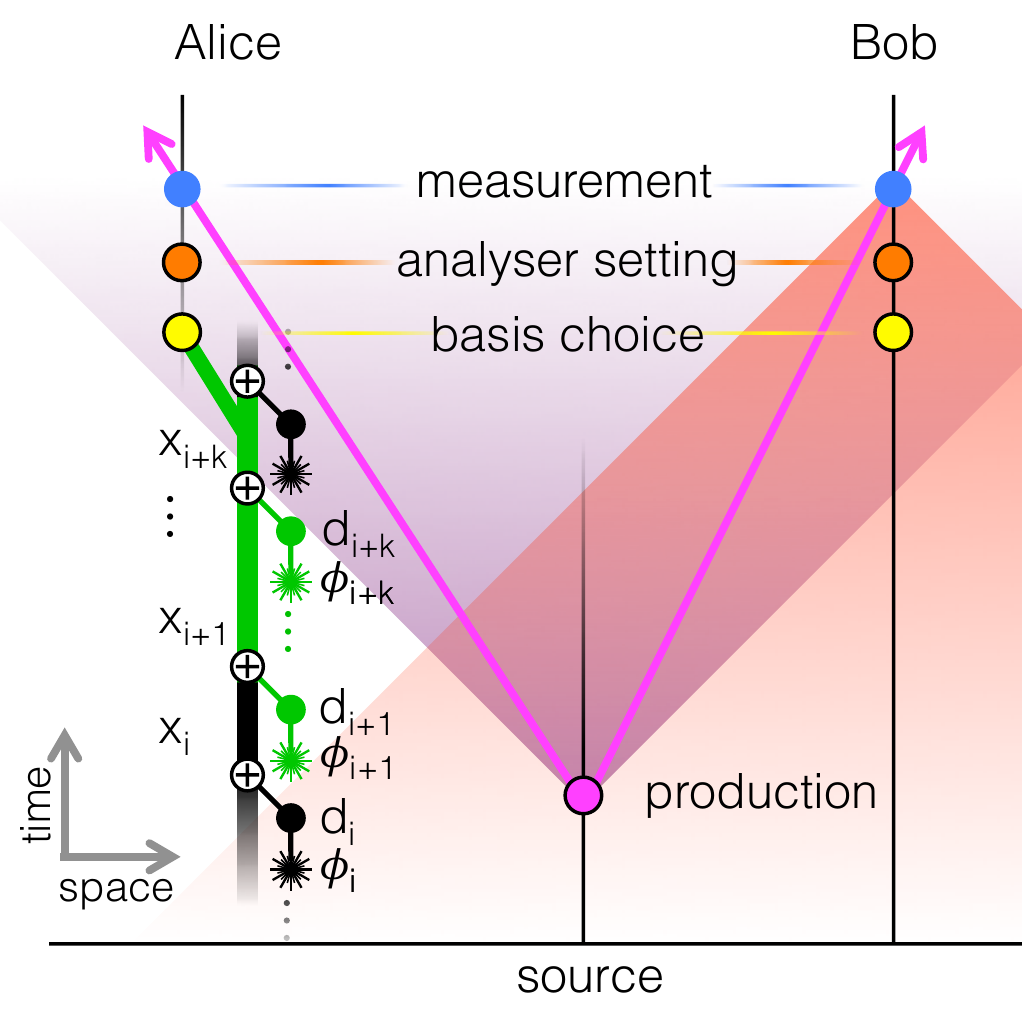} 
\caption{\label{Fig:SpaceTimeDiagram} Space-time diagram for the production of random numbers in a loophole-free Bell scenario. As shown, up to $k$ raw bits can be generated in a time window that is spacelike separated from both (i) the pair generation and (ii) the distant measurement.  Laser pulses (\Laserbeam) with random phases $\phi_i$ are converted into raw random bits $d_i$ and extracted bits $x_i$ by a running \xor~$(\oplus)$ calculation. }
\end{figure}

Time and/or frequency metrology, e.g.  jitter measurements against stabilized oscillators, are routinely used to determine  timing with sub-ns precision and accuracy, allowing reliable identification of spacelike separated events.  Achieving similar assurances for unpredictability poses a distinct challenge.  For fundamental reasons, no test on the {output} of a RNG can demonstrate randomness, and statistical characterization of the RNG {process} becomes the critical task.  Here we develop statistical metrology for a short-delay RNG, in analogy to earlier work with high-throughput RNGs \cite{XuOE2012, AbellanOE2014, MitchellPRA2015}. The {\prederr} $\epsilon$ is exponentially reduced by randomness extraction  %,Frauchiger:2013tf
(RE) \cite{VadhanFTTCS}:  in real time, we compute the parity of several raw bits to produce one {very unpredictable} extracted bit for the setting choice. 

\begin{figure}[b]
\centering
\includegraphics[width=0.8\linewidth]{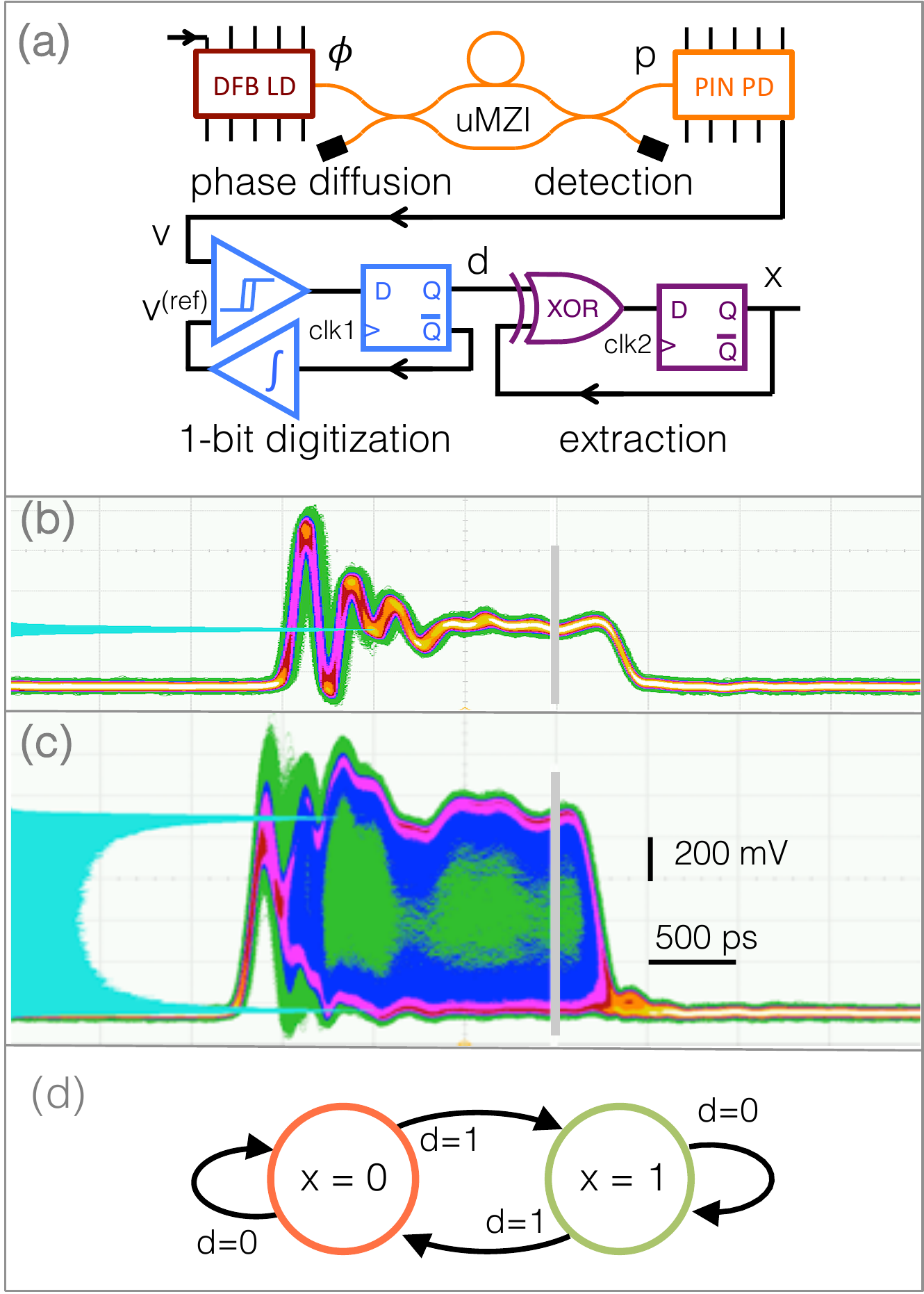} 
\caption{\label{Fig:setupMain} Random number generation for loophole-free Bell tests. (a) Experimental schematic. Laser pulses 
%(\Laserbeam) 
with random phases $\phi_i$ from a distributed feedback laser diode (DFB LD) are converted to random powers $p_i$ by an UMZI and detected with a linear photoreceiver (PIN PD) to give analog voltages $v_i$.  These are one-bit digitized with a comparator and D-type flip-flop to give raw bits $d_i$, and summed {modulo 2} %($\oplus$)
 {with an \xor~ gate} to give extracted bits $x_i$.   The output value $x_{i+k}$ includes the parity of $k$ raw bits, $d_{i+1}$ to $d_{i+k}$,  due to pulses spacelike separated from the distant measurement and from the entanglement production.   (b), (c) ``Persistence mode'' visualization of $v(t)$ statistics. Warmer colors show greater frequency; teal histogram on the left axis describes the voltages sampled inside {the time window indicated in gray.}  (b)  Noninterfering pulses obtained by blocking the long interferometer path; (c)  interfering pulses.  (d)  Two-state machine describing the randomness extraction.}
\end{figure}

{The {RNG} and its behavior are illustrated  in Fig.~\ref{Fig:setupMain}. A single-mode {laser diode} ({LD}) is strongly current modulated, going above threshold for about 2 ns of every 5 ns cycle, to produce a train of optical pulses with very similar waveforms, as seen in Fig.~\ref{Fig:setupMain}~({b}). In the time below threshold strong phase diffusion randomizes the optical phase within the laser \cite{Henry:1982dv,NieRSI2015}, and, thus, the relative phase $\phi$ from one pulse to the next. At the time a pulse leaves the laser, it is already a macroscopic ($\sim$ mW) signal, with a phase that has been fully randomized by the microscopic process of spontaneous emission.  An unbalanced Mach-Zehnder interferometer (UMZI) converts the train of phase-random pulses into amplitude-random pulses; see Fig.~\ref{Fig:setupMain}~(c).  These are detected with a fast photodiode, giving a voltage signal $v(t)$.  A fast comparator and a D-type flip-flop digitize (with one-bit resolution) the signal at times $t_i$ to give, {at \SI{200}{Mbps}}, raw digital values $d_i = \theta\textbf{(}v(t_i) - \vref(t_i)]\textbf{)}$, where $\theta$ is the Heaviside step function and $\vref$ is the comparator reference level.  To correct for drifts in laser power, the reference level is set by feedback from the raw digital values via an integrator with a 1 ms time constant.  We observe a raw-bit average of 
$\expect{d} = \frac{1}{2}[1 + 6.9(1)
 \times 10^{-4}]$.

 An \xor~gate and a second flip-flop perform a running parity calculation, updating the output $x$ as $x_i = x_{i-1} \oplus d_i$, where $\oplus$ indicates addition {modulo 2}. This describes a two-state machine, see Fig.~\ref{Fig:setupMain}~(d), that changes state every time a new raw bit $d_i=1$. 
Note that $x$ accumulates the parity of all preceding raw bits, only $k$ of which will be spacelike separated from the distant measurement.  When a bit $x_{i+k} = x_i \oplus d_{i+1} \oplus \ldots \oplus d_{i+k}$ is used for a basis setting, $x_i$ contributes no spacelike separated randomness, and the predictability of $x_{i+k}$ will be determined by $d_{i+1} \oplus \ldots \oplus d_{i+k} \equiv \dik $.} Writing the predictability of $d_i$, i.e. the probability of the more likely value, as $\Pred(d_i) = \frac{1}{2} ( 1 + \epsilon_i)$, where $\epsilon_i \ge 0$ is the instantaneous \prederr, we find (see the Supplemental Material) that if $\epsilon_i \le\epsilon_{\rm max}$, the predictability of the parity of $k$ bits is bounded as $\Pred(\dik) \le  \frac{1}{2} ( 1 + \epsilon_{\rm max}^k)$. The RE output approaches ideal randomness exponentially in $k$.  

\newcommand{\barA}{\overline{A}}

{
We define the ``freshness time'' to be the interval between the earliest spontaneous-emission events required for randomizing a bit and the bit's availability for use. The largest phase diffusion occurs at the rising edge of the current pulse, when the intracavity photon number is {at a} minimum \cite{Henry:1982dv}. {The freshness time for a single bit $\ftime$, measured from a rising edge of the electrical modulation signal to availability of the corresponding bit at the output port is bounded by $10.01<\ftime<11.07$ ns with a p-value $<  1.4 \times 10^{-6}$ (see the Supplementary Material). Since we can use $(k-1)$ extra bits that are still spacelike separated, $(k-1)$ additional clock cycles of \SI{5} ns each are needed. In total, the freshness time to produce and propagate $k$ bits from the oldest spacelike separated spontaneous-emission event to the output port is $\ftime^{(k)} = \ftime + 5\times(k-1)$. }

}
\begin{figure}[t]
\centering
\includegraphics[width=0.80\linewidth]{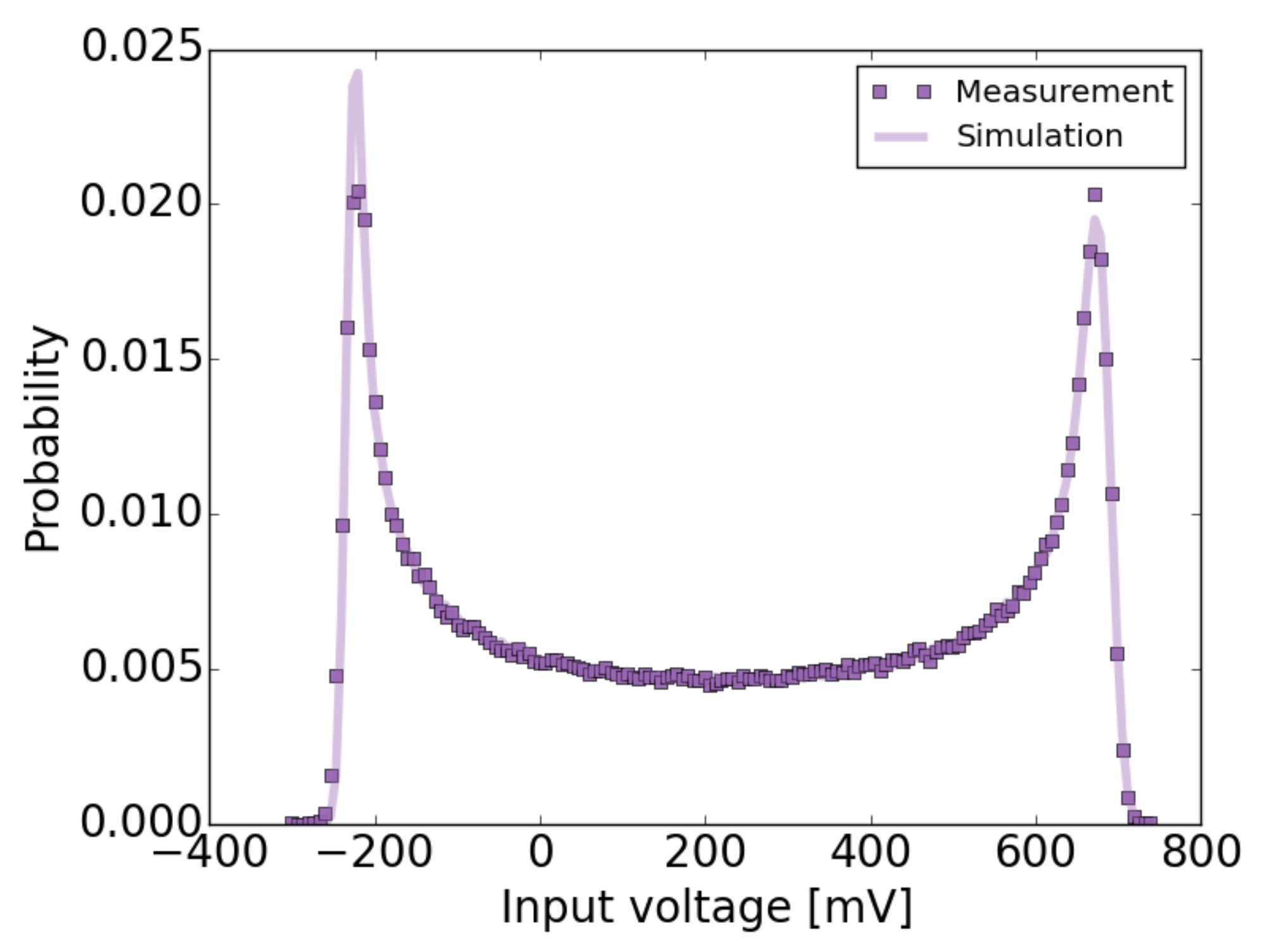} 
\caption{\label{Fig:Histogram}Histogram (points) of analog signals $v$, showing an arcsine distribution, and prediction (line) from a Monte Carlo simulation of Eq. (\ref{Eq:2}) using measured rms deviations for all noise sources ${\vS}, {\vL}, {\vD}, {\vhang}$ and ${\vref}$, and a fitted visibility ${\cal V} = 0.955$.  Voltage scale is offset due to ac-coupling. }
\end{figure}

{
Metrological assurances proceed from the interference behavior.  The instantaneous power reaching the detector is
\be
p_I(t) = \pS(t) +  \pL(t) +  2 \sqrt{\pS(t) \pL(t)} \cos \phi(t),
\ee
where $\pS$ and $\pL$ are the contributions of the short and long paths, respectively. We note that optical visibility is guaranteed by the single-spatial-mode fiber interferometer and the single-longitudinal-mode laser emission. 
Including detection noise and finite bandwidth effects, the electronic output is (see the Supplemental Material)
\bea
\label{Eq:2}
v(t) & = &   \vS(t) + \vL(t) +  \vhang(t) + \vD(t) + \vphi(t),
\eea
where $\vS$ and $\vL$ are the short- and long-path contributions, respectively, $\vhang$ describes ``hangover errors,'' i.e. delayed contributions from earlier pulses \cite{MitchellPRA2015}, and $\vD$ is the detector noise.  $\vphi = 2\vis \sqrt{\vS \vL} \cos \phi$ is the trusted signal from interference, where $\vis$ is the visibility after detection. In Fig.~\ref{Fig:Histogram}), we show the distribution of $v$, sampled at the moment indicated in Fig.~\ref{Fig:setupMain}~(c), and infer $\vis\sim 95$\% using a Monte Carlo simulation of Eq. (\ref{Eq:2}) (see the Supplementary Material). As shown in Fig.~\ref{Fig:setupMain}~{(c)}, we take a sample $<\SI{2}{ns}$ after the rising edge occurs, chosen late in the pulse so that relaxation oscillations have decayed. The histogram is well modeled by the arcsine distribution, which describes the cosine of a uniformly distributed phase. 

{
The trusted randomness of the signal $v$ originates in $\phi$, which between pulses strongly diffuses due to spontaneous emission, as shown in Fig.~\ref{Fig:diffusive} (see, also, the Supplemental Material).  The  observable \mbox{$\phi \!\! \mod 2\pi$} is for all practical purposes uniformly distributed on $[0,2\pi)$, is unpredictable based on prior conditions, and is independent from one pulse to the next \footnote{This applies not only in a  quantum description of laser operation, but in any theory with unpredictable laser phase diffusion. See Supplementary Materials.}, irrespective of any other phase shifts \cite{MitchellPRA2015}.
}

\begin{figure}[t]
\centering
\vspace{0.2cm}
\includegraphics[width=0.72\linewidth]{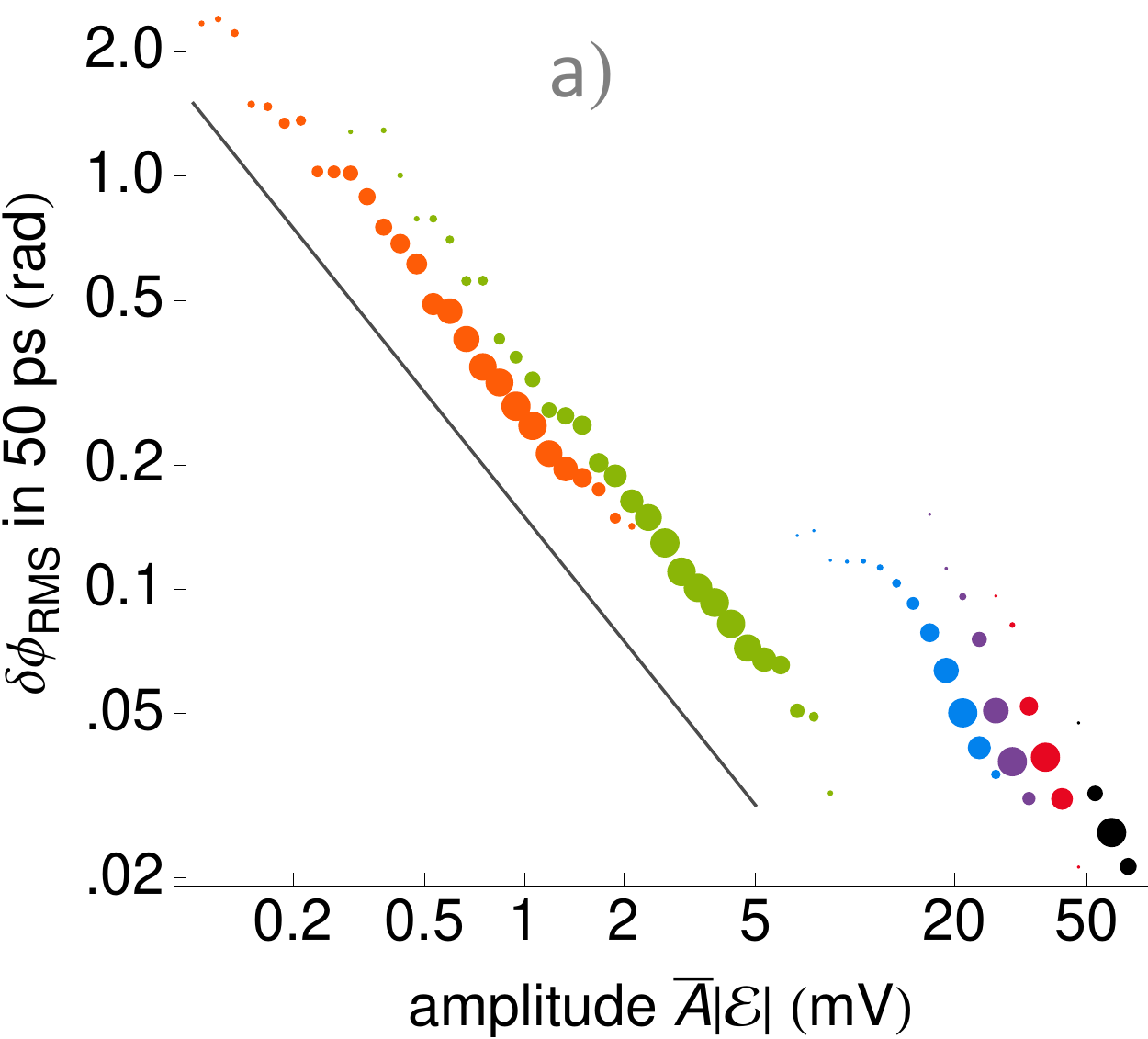} 
\caption{\label{Fig:diffusive} Observed frequencies (spot size) of field amplitude $|{\cal E}|$ and resulting phase dispersion $\delta \phi_{\rm rms}$ over {50}{ ps}, measured by heterodyne detection with gain $\barA$ for continuous currents (colors, left to right) 15, 16, 16.5, 17, 17.5, {19}{ mA}.  Grey line shows  $\delta \phi_{\rm rms} \propto |{\cal E}|^{-1}$ scaling of spontaneous-emission-driven phase diffusion. }
\end{figure}

With the exception of $\cos\phi$, all contributions to $v(t)$ in Eq. (\ref{Eq:2}), and also $\vref$, contain technical noise due to prior conditions that are not spacelike separated from the distant detection. We define the sum of these untrusted contributions $\vc \equiv \vS +\vL + \vD + \vhang - \vref $  so that  $d_i = \theta[ v(t_i) - \vref(t_i)] =  \theta[ \vphi(t_i) + \vc(t_i)]$, with distribution 
\bea
\label{eq:PredFromVc}
P(d=1) &\equiv& P_1 =  \frac{2}{\pi} \arcsin \sqrt{\frac{1}{2} + \frac{\vc}{2\vrange}} 
\eea
where $2 \vrange = 4 \vis \sqrt{\vS \vL}$ is the peak-to-peak range of $\vphi$ (see the Supplemental Material).  The predictability is $\Pred(d) \equiv \max [ P_1, 1-P_1 ]  \equiv \frac{1}{2}(1+\epsilon)$. Bounding the effect of $\vc$ on $\Pred(d)$ will determine $\epsilon_{\rm max}$, the upper bound on $\epsilon$.    
}

{The contributors to $\vc$ are electronic signals and are directly measured with a {4 GHz} oscilloscope {(Agilent Infinitum MSO9404A)}. For example, the variation of $\vS$, the signal in the short path of the interferometer, is measured by blocking the signal from the long path. The measurement gives access to the signal $\vS+\vD+\vO$, as shown in Fig.~\ref{Fig:setupMain}~(b). Note that the measurement of the signal of the short path is not isolated, but superimposed to the noises in the photodetector $\vD$ and the scope $\vO$. To obtain the noise from $\vS$ only, we have to subtract the contribution from $\vD$ and $\vO$, both directly measurable. Statistics of the measurable noise contributions, always sampled at the same point in the pulse, are given in Table~S1 of the Supplemental Material.  
}

 \newcommand{\ordinary}{ordinary} 
\newcommand{\paranoid}{digitizer paranoid} 
\newcommand{\extreme}{fully paranoid} 

{
To combine the noise sources, we consider three levels of distrust of the equipment: ``\ordinary,'' ``\paranoid'' and ``\extreme.''  In all cases, the noises are individually described by the measured statistics of Table~S1 in the Supplemental Material, but their assumed correlations vary.  In ordinary distrust, we make the physically reasonable assumption that the noise sources are uncorrelated.   In digitizer paranoid distrust we assume the comparator, the only nonlinear element of the signal chain, chooses $\vref$ in function of the other noises so as to maximize the predictability. In fully paranoid distrust, we assume that all noise sources are collaborating to maximize predictability. 
These assumptions lead to normally-distributed $\vc$ with rms deviations $\sigma$ shown in Table \ref{Table:untrusted}.  Fluctuations in $\vc$ are, in principle, unbounded but rarely exceed a few standard deviations, a situation that is captured by assigning confidence bounds, in this case to $\Pred(d)$ and $\Pred(x)$.  For example, considering $\vc = | \vcbar | + 6\sigma$ as an upper limit, we compute $\epsilon_{\rm max}$ using Eq. (\ref{eq:PredFromVc}). 
{Noise fluctuations will produce a fraction $P_{6\sigma}$ of the raw bits with $\epsilon >\epsilon_{\rm max}$, where $P_{6\sigma} \approx 2 \times 10^{-9}$.}  The {\prederr} of the extracted bit exceeds $\epsilon_{\rm max}^k$ at most this often, even assuming maximally correlated raw-bit \prederr. 
See the Supplemental Material for details. Values of $\sigma_{\vc}$, $\epsilon_{\rm max}$ and $\ftime$ for different $k$ and distrust levels are shown in Table \ref{Table:untrusted}, and in Fig.~\ref{Fig:xcorr}. 
}

\begin{table}[!htp]
\caption{\label{Table:untrusted} Noise and  predictability for different trust scenarios.
All predictabilities are given for a $6\sigma$ confidence level. Times in parentheses indicate {freshness time $\ftime^{(k)}$.}  %Production times $\ptime$ are shorter by at least \SI{8}{ns}.}
See text for details.  }
\centering
\begin{tabular*}{1\linewidth}{ L{0.17\linewidth} C{0.17\linewidth} C{0.30\linewidth} C{0.3\linewidth} C{0 \linewidth}}
\toprule
Distrust level & Noise $\sigma_{\vc}$ & {\Prederr} \mbox{$\epsilon_{\rm max}^4$} (26 ns) & {\Prederr} $\epsilon_{\rm max}^6$ (36 ns)&  \rule[0\baselineskip]{0pt}{1.1\baselineskip} \\
\hline
Ordinary & $8.6$ mV  & $2.5\times 10^{-5}$ &$1.3\times 10^{-7}$ & \rule[0\baselineskip]{0pt}{1.1\baselineskip}  \\
Dig.~par.&  $11.7$ mV & $8.6 \times 10^{-5}$   & $8.0\times 10^{-7}$ \\
Fully~par.& $14.5$ mV &$2.0\times 10^{-4}$  & $2.9\times 10^{-6}$ \\
\toprule
\end{tabular*}
\end{table}

{
Although no test of the output can assure randomness, tests can, nonetheless, detect failure to be random. Because of the low computational capacity of physical RNGs, imperfections are expected mostly in  low-order correlations.  The autocorrelation of the extracted output {$\Gamma_x(k) \equiv \langle x_{i} x_{i+k} \rangle - \langle x_{i}\rangle^2$} is bounded by $4|\Gamma_x(k)| \le \epsilon_{\rm max}^{k}$ and, thus, drops off in the same way as the {\prederr} (see the Supplemental Material).   As shown in Fig.~\ref{Fig:xcorr}, the measured $|\Gamma_x(k)|$  approaches zero exponentially in $k${, and already with $k=4$ reaches $|\Gamma_x(k)|< 10^{-6}$, the statistical limit with 1 Tbit. }
}

\newcommand{\mysc}[1]{\textsc{\small #1}}

{As detailed in the Supplemental Material, we have applied statistical tests \mysc{DIEHARDER} \cite{BrownWEB2004b}, \mysc{NIST SP800-22} \cite{RukhinNIST2010tech}, and  \mysc{TESTU01} \mysc{ALPHABIT} battery \cite{LEcuyerACM2007} to strings of extracted data up to \SI{1}{Tbit} in length.  To study a given $k$, we first generate a distilled string $z_i \equiv x_{i k}$, i.e. $\{z\}$ is a $k$-fold subsampling of $\{x\}$.  
We observe that \mysc{ALPHABIT}, designed to test physical RNGs, is as sensitive as other tests and runs much faster.  {Already with $k=3$ extraction \mysc{ALPHABIT} finds no significant patterns in $\sim$ 2.3 Tbits of data, organised as one file of 1 Tbit, two files of 500 Gbits, one file of 80 Gbits, and two files of 64 Gbits. We also tested 300 sequences of lengths 1 Mbit, 0.2 Gbits, 0.5 Gbits, 1.0 Gbit for $k= 1,2,3,$ and $4$, respectively, and compare the failure rates to what is expected for an ideal random source.}}

\begin{figure}[t]
\centering
\vspace{0.2cm}
\includegraphics[width=1\linewidth]{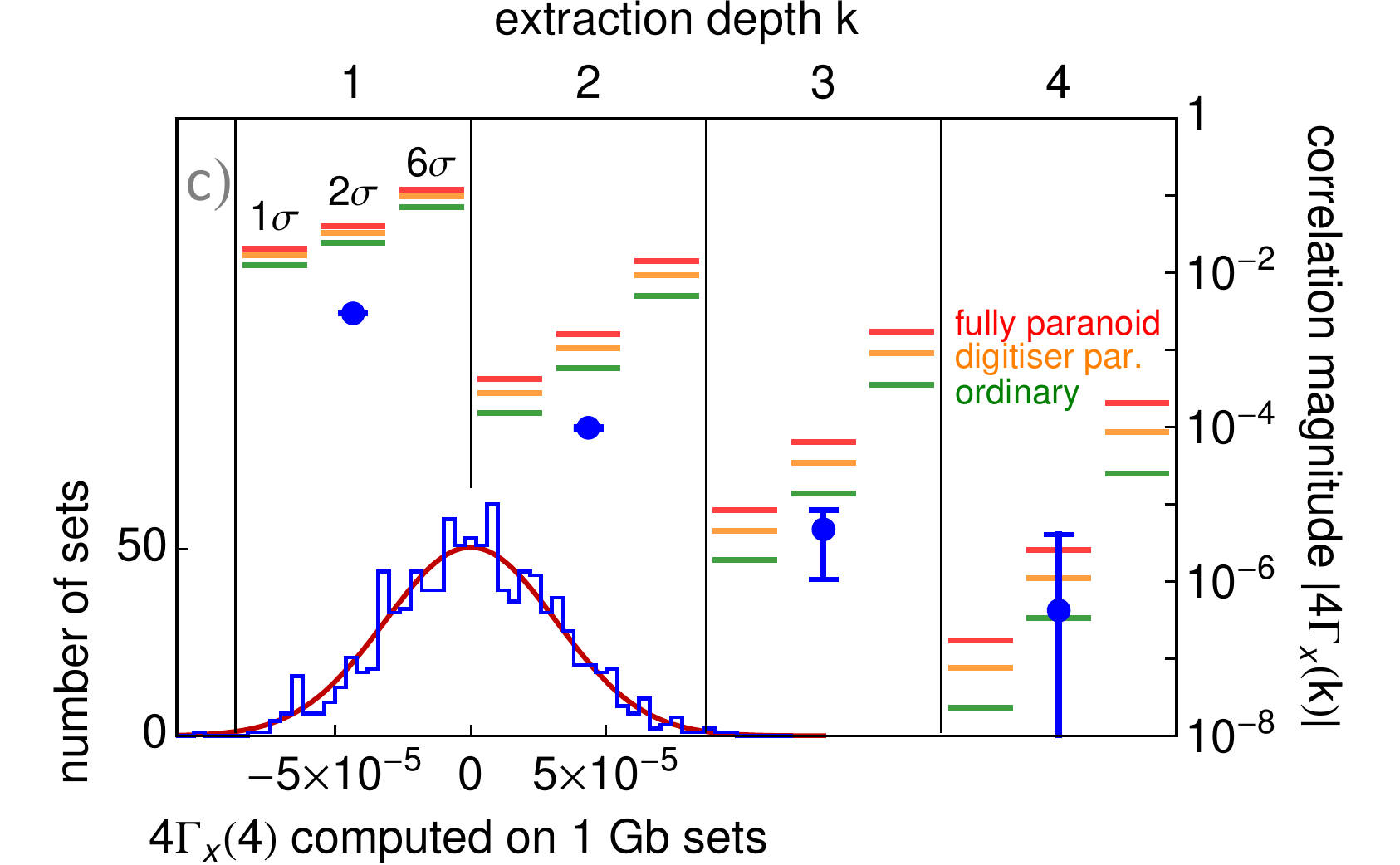} 
\caption{\label{Fig:xcorr}Measured two-point correlations (blue points) $|4 \Gamma_x(k)|$ for extracted bits $x$, computed on 300 Gbits, showing exponential approach to ideal behavior.  Error bars show $\pm 1 \sigma$  statistical error in the correlation measurement.  Horizontal bars show $\epsilon_{\rm max}^k$, the metrologically derived upper bound on predictability, for ordinary, (green) digitizer paranoid (orange) and fully paranoid (red) distrust levels and (left to right)  $1 \sigma$, $2 \sigma$, and $6 \sigma$ confidence. { (Lower left) Histogram of $4\Gamma_x(4)$ computed on 1000 runs of \SI{1}{Gb} each, is normal by the Pearson $\chi^2$ test  ($p=0.27$).  Curve shows ideal distribution.     } }
\end{figure}
  
In conclusion, we have demonstrated a spontaneous-emission-driven random number generator suitable for closing the locality and freedom-of-choice loopholes in a test that also closes the detection loophole. By combining high-speed phase-diffusion RNG, real-time randomness extraction, and metrological guarantees, we have produced extracted bits traceable to spontaneous-emission events less than \SI{36}{ns} old and with {\prederr} $\epsilon \le 10^{-5}$.  Generation of  high-quality random bits in narrow time windows enables definitive tests of quantum nonlocality and ``device-independent'' technologies guaranteed by the no-signaling principle.

%[1,500 words of text (excluding the first paragraph of Letters, figure legends, reference list and the methods section if applicable) ]

{

{Acknowledgements:}
We thank  M. Giustina, B. Hensen, K. Shalm, J. Kofler,  S. Wehner,  S. Glancy, S. Jordan, M. Wayne, J. Bienfang, R. Mirin, C. Marquardt, R. Hanson, S.-W. Nam and A. Zeilinger
for helpful discussions, and the ICFO electronic workshop for peerless craftsmanship.  We thank F. A. C. Diaz-Balart and F. A. C. Smirnov for particularly stimulating conversations. 
The work was supported by the European Research Council project AQUMET, FET Proactive project QUIC,  Spanish MINECO projects MAGO (Ref. FIS2011-23520) and EPEC (FIS2014-62181-EXP), Catalan 2014 SGR 1295, the European Regional Development Fund (FEDER)  grant TEC2013-46168-R,  and by Fundaci\'{o} Privada CELLEX.  

%{\color{white} 
%\footnote{See Supplemental Material attached to this document, which includes Refs. [36-42].}
% \cite{AgrawalJQE1990, ScullyZubairy1997}
% \cite{RauchAIAAJ1965,GelbBook1974}
% \cite{EinsteinDPG1916}
% \cite{LarssonJPA2014}
%  \cite{JacobsenMaster2014}
%  }

%\input{./FQRNG151015main.bbl}
\bibliographystyle{apsrev4-1no-url}
\bibliography{./freshqrng}

%[MAX 30 references !]

\clearpage

%%%%%%%%%%%%%%%%%%%%%%%%%%%%%%%%%%%%%%%%%%%%%%%%%%%%%%%%%%%%%%%%%%%%%%%%%%%%%%%%%%%%%%%%%%%%%%%%%%%%%%%%%%%%%%%%%%%%%%%%%%%%%%%%%%%%%%%%%%%%%%%%%%%%%%%%%%%%%%%%%%%%%%%%%%%%%%%%%%%%%%%%%%%%%%%%%%%%%%%%%%%%%%%%%%%%%%%%%%%%%%%%%%%%%%%%%%%%%%%%%%%%%%%%%%%%%%%%%%%%%%%%%%%%%%%%%%%%%%%%%%%%%%%%%%%%%%%%%%%%%%%%%%%%%%%%%%%%%%%%%%%%%%%%%%%%%%%%%%%%%%%%%%%%%%%%%%%%%%%%%%%%%%%%%%%%%%%%%%%%%%%%%%%%%%%%%%%%%%%%%%%%%%%%%%%%%%%%%%%%%%%%%%%%%%%%%%%%%%%%%%%%%%%%%%%%%%%%%%%%%%%%%%%%%%%%%%%%%%%%%%%%%%%%%%%%%%%%%%%%%%%%%%%%%%%%%%%%%%%%%%%%%%%%%%%%%%%%%%%%%%%%%%%%%%%%%%%%%%%%%%%%%%%%%%%%%%%%%%%%%%%%%%%%%%%%%%%%%%%%%%%%%%%%%%%%

\begin{widetext}
%Quantum random numbers with fast randomness extraction for nanoerrors in nanoseconds  \\
%Real-time randomness extraction for nanoerror  random bits on nanosecond budgets \\
%Nanosecond  extraction of  quantum randomness.
%Real-time extraction of quantum randomness: exponential error reduction in nanoseconds\\
{\Large Supplemental material for ``Generation of fresh and pure random numbers for loophole-free Bell tests''} \\
%\btext{Real-time extraction of bounded quantum randomness:  nanoerrors in nanoseconds}
%\btext{Real-time extraction of spontaneous emission randomness:  nanoerrors in nanoseconds}
%\rtext{CONFIDENTIAL : DO NOT DISTRIBUTE} \\
%{Generation of fresh and pure random numbers for loophole-free Bell tests}
%\\{\small 90 characters, max, including spaces}
\end{widetext}

\author{Carlos Abellan}
\affiliation{ICFO -- Institut de Ciencies Fotoniques, The Barcelona Institute of Science and Technology, 08860 Castelldefels (Barcelona), Spain}

 \author{Waldimar Amaya}
\affiliation{ICFO -- Institut de Ciencies Fotoniques, The Barcelona Institute of Science and Technology, 08860 Castelldefels (Barcelona), Spain}

\author{Daniel Mitrani}
\affiliation{ICFO -- Institut de Ciencies Fotoniques, The Barcelona Institute of Science and Technology, 08860 Castelldefels (Barcelona), Spain}

\author{Valerio Pruneri}
\affiliation{ICFO -- Institut de Ciencies Fotoniques, The Barcelona Institute of Science and Technology, 08860 Castelldefels (Barcelona), Spain}
\affiliation{ICREA -- Instituci\'{o} Catalana de Recerca i Estudis Avan\c{c}ats, 08015 Barcelona, Spain}

\author{Morgan W. Mitchell}
\affiliation{ICFO -- Institut de Ciencies Fotoniques, The Barcelona Institute of Science and Technology, 08860 Castelldefels (Barcelona), Spain}
\affiliation{ICREA -- Instituci\'{o} Catalana de Recerca i Estudis Avan\c{c}ats, 08015 Barcelona, Spain}

\date{\today}

%\begin{abstract}
%\end{abstract}

\renewcommand{\omit}[1]{}

\maketitle

\renewcommand{\theequation}{S\arabic{equation}}
\setcounter{equation}{0}
\renewcommand{\thetable}{S-\Roman{table}}
\setcounter{table}{0}
\renewcommand{\thefigure}{S\arabic{figure}}
\setcounter{figure}{0}
%\renewcommand{\bibnumfmt}[1]{[S#1]}
%\renewcommand{\citenumfont}[1]{S#1}

%\section{supplemental material}
%\section{Theory}

%\scriptsize \small
%\subsection*{METHODS} %(optional)

%\cite{AgrawalJQE1990, ScullyZubairy1997}
%% \cite{AbellanOE2014,Yuan:2014co}
% \cite{RauchAIAAJ1965,GelbBook1974}
% \cite{EinsteinDPG1916}
%% \cite{KoflerARX2014}
% \cite{LarssonJPA2014}
%%  \cite{RukhinNIST2010tech}
%  \cite{JacobsenMaster2014}

\section{Device construction} 
The optoelectronic components (LD and PD) are commercial telecommunications devices designed for $\ge$ \SI{10}{Gbps} direct-modulation data transmission at 1550 nm.  The LD is temperature controlled with a Peltier element to ensure wavelength stability and incorporates an optical isolator to prevent optical feedback.  The PD incorporates a low-noise trans-impedance amplifier with a linear response.  The interferometer is built of polarisation-maintaining single-mode fibre with fibre lengths cut to make the delay of the uMZI \SI{5}{ns}, equal to the pulse repetition period.  The logic elements (comparator, flip-flops and exclusive-OR) are designed for $\ge$ \SI{10}{GHz} operation with $\le$ \SI{5}{ps} jitter.  A programmable clock distribution integrated circuit is used to adjust timings with \SI{10}{ps} resolution to ensure the relative timing of sampling and logic operations.  Printed circuit boards were designed, assembled and tested by the ICFO electronic workshop.  

\section{Spontaneous emission driven phase diffusion}
 Laser phase diffusion ({LPD}) has been intensively studied in {LD}s, where it is responsible for the free-running line-width \cite{Henry:1982dv}.  {LPD} in {LD}s is driven by spontaneous emission, spontaneous carrier recombination, {Johnson noise in the current supply and technical noise sources such as environment-induced current fluctuations.}  Spontaneous emission contributes a delta-correlated Langevin force \citeSOM{AgrawalJQE1990, ScullyZubairy1997}, which when integrated makes a  contribution to $\phi$ that is independent from one pulse to the next. In the \SI{5}{ns} period considered here, this contribution is  sufficiently large that the distribution of $\phi$ wraps several times around the phase circle \cite{AbellanOE2014,Yuan:2014co}. %{Heterodyne measurements for Fig.~\ref{Fig:Validation}~a) are described in the supplemental material.} 

To validate this model of phase diffusion, we perform heterodyne detection, beating the tested laser against a second ``local oscillator'' (LO) laser, both running at constant injection current. The beat note, tuned to $\sim$ \SI{3}{GHz} by temperature adjustment of the LO, is detected with a fast photodiode (ThorLabs DET08CFC, \SI{5}{GHz} bandwidth), and digitized on a fast oscilloscope (Keysight/Agilent Infiniium MSO9404A) with bandwidth \SI{4}{GHz} and sampling rate \SI{20}{GSa/s}.  The observed signal is
\bea
v(t) &\propto& |  {\cal E}_{\rm LO} (t) +  {\cal E}(t) |^2 
\nne
|  {\cal E}_{\rm LO} (t)|^2 +   {\cal E}_{\rm LO}(t) {\cal E}^*_{}(t) + {\cal E}^*_{\rm LO}(t) {\cal E}_{}(t) +   |{\cal E}(t) |^2 \nonumber \\
 &=& \barA^2 + 2 \delta A(t) + \barA\left[ \cos \Omega t \,{\rm Re}\, {\cal E}(t) + \sin \Omega t \, {\rm Im}\, {\cal E}(t) \right] \nnp O(\delta A^2) + O(\delta A |{\cal E}|) + O(|{\cal E}|^2),
\eea
where the last three terms are small.  Here $A(t) \equiv \barA + \delta A(t) \equiv |{\cal E}_{\rm LO} (t)|$ describes the slowly-varying LO strength, $\Omega$ is the angular frequency of the LO in a frame rotating at the mean test laser frequency, and ${\cal E}(t) = |{\cal E}(t)| \exp[i \phi(t)]$ is the test laser field, also in the rotating frame.  Linearising the model by dropping the last three terms, we extract $\delta A(t), {\rm Re}\, {\cal E}(t)$ and ${\rm Im}\, {\cal E}(t)$ by Rauch-Tung-Striebel smoothing, i.e., by bi-directional Kalman filtering \citeSOM{RauchAIAAJ1965,GelbBook1974}. 

%The results, shown in Fig. \ref{Fig:Holevo}, indicate both a diffusive phase behaviour, i.e. a linear increase of $\holvar \Delta\phi$ with $\Delta t$, and a rapid increase of the diffusion rate as the current drops below threshold. Already at \SI{16}{mA}, or $\approx$ 80\% of threshold, the diffusion is by several $\pi$ in \SI{1}{ns}.  This direct observation, along with the fact that in the pulsed sequence the current remains far below threshold for  $\approx$ \SI{3}{ns}, 
% supports the theoretical expectation \citeSOM{Henry:1982dv,AbellanOE2014} that the phase diffuses by $\delta \phi \gg \pi$ and thus for all practical purposes is completely randomized. 
\begin{figure*}[t]
\centering
\includegraphics[width=0.86 \textwidth]{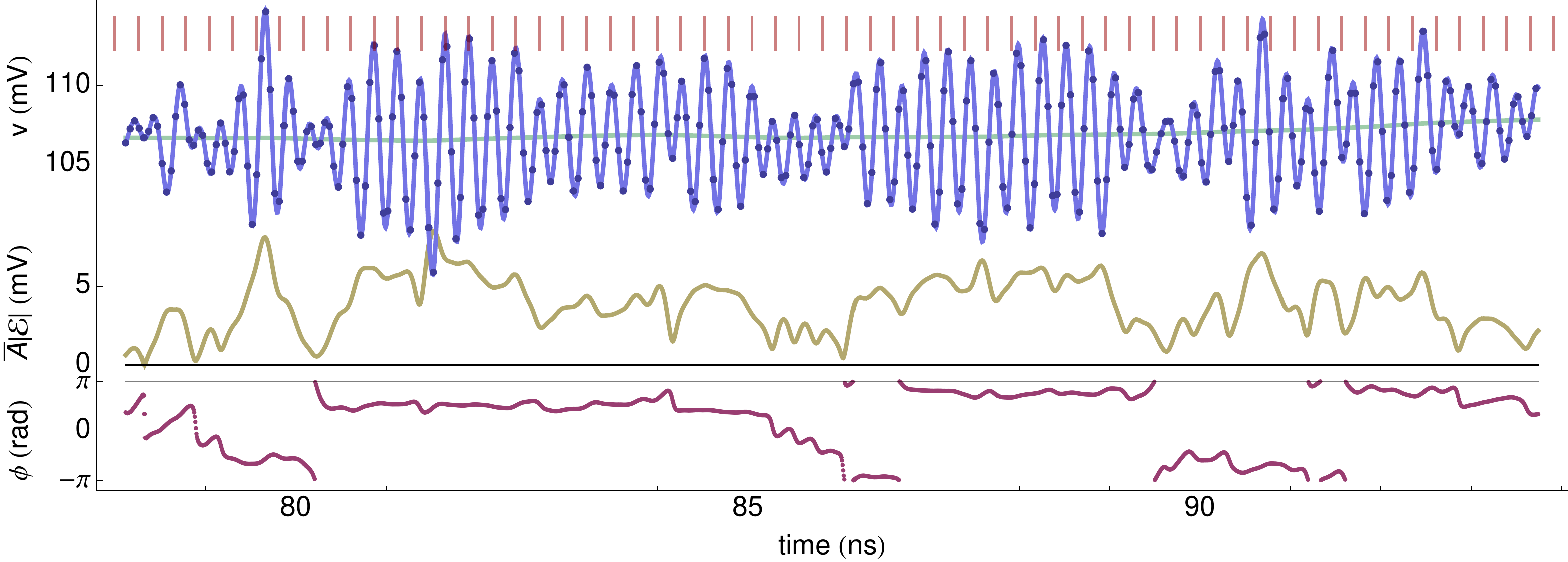}
\caption{{Direct observation of spontaneous emission driven phase diffusion in the DFB laser at a constant, threshold-region drive current of \SI{18}{mA}.  Heterodyne signal $v(t)$ (top, blue): Points show \SI{20}{Gsa/s} oscilloscope data, curve shows prediction of Kalman filter, from which amplitude $\barA|{\cal E}|$ (middle, beige), phase $\phi$ (bottom, maroon) and baseline   $A^2$  (top, green) are obtained. Ruler (top, red) is a guide to make phase changes more visible.  As expected for phase diffusion by spontaneous emission, rapid phase changes occur when the amplitude dips toward zero.    } }
\label{Fig:HeterodyneData} 
\end{figure*}

As seen in  Fig. \ref{Fig:HeterodyneData}, the signals show direct evidence for phase-diffusion by spontaneous emission: at large or moderate amplitudes, the phase is stable for several cycles.  It makes rapid changes, however, either increasing or decreasing the phase, when the amplitude $|{\cal E}|$ becomes small.  In contrast to a laser above threshold, which diffuses in phase but returns to a finite equilibrium amplitude, in a laser below threshold ${\cal E}$ executes a diffusive motion about the point ${\cal E} = 0$.  Because they contribute to the field with random phases, spontaneous emission events drive diffusion equally in the real and imaginary components of ${\cal E}$, giving a phase diffusion coefficient inversely proportional to the amplitude \citeSOM{AgrawalJQE1990, ScullyZubairy1997}.  This behavior is quantified in Fig.~4, which shows statistics acquired from \SI{800}{ns}-long traces of signals of the kind shown in Fig. \ref{Fig:HeterodyneData}.  

We quantify the phase diffusion using the Holevo variance $\holvar(\phi) \equiv |\langle \exp[i \phi] \rangle|^{-2} -1$, which is well-behaved for cyclic variables and approaches the ordinary variance $\var (\phi) \equiv \langle \phi^2 \rangle - \langle \phi \rangle^2$ for small values.  For the data shown in Fig.~4, we choose a time separation $\Delta t = \SI{50}{ps}$, corresponding to the oscilloscope sampling period, and compute the phase diffusion as 
\bea
\delta \phi_{\rm RMS}
%({|{\cal E}|}) 
&\equiv& \sqrt{ |\langle e^{i [\phi(t+\Delta t) - \phi(t) ] } \rangle_{|{\cal E}|} |^{-2} -1 },
\eea
where the angle brackets $\langle \cdot \rangle_{|{\cal E}|}$ indicate the average over all times $t$ with a given $|{\cal E}|$.  
%The reported $\delta\phi_{\rm RMS}$ is computed as the square root of $\holvar [\phi(t + \SI{50}{ps}) - \phi(t)]$, this time on the 
The scaling $\delta \phi_{\rm RMS} \propto |{\cal E}|^{-1}$ observed below threshold is a direct confirmation that spontaneous emission is responsible for the observed strong phase diffusion.  Phase changes due to refractive index variation would cause $|{\cal E}|$-independent phase changes and fluctuating nonlinearities would give  diffusion increasing with $|{\cal E}|$.  

The \SI{18}{mA} current used in Fig. \ref{Fig:HeterodyneData} is near the lasing threshold, resulting in multi-ns periods of slow diffusion.  In contrast, when used in the RNG, the laser is taken far below threshold to produce much faster phase diffusion and maintained in this condition for \SI{3}{ns} to achieve a full  spontaneous emission driven phase randomization.

{
\section{Unpredictability of phase diffusion in non-quantum theories}

Bell inequalities test local realistic hidden-variable theories (HVTs), and must be analyzed under the assumptions of local realism, not those of quantum mechanics.  Spontaneous emission and laser phase diffusion are observable phenomena and unlike, say, entanglement, do not in themselves belong to any particular theory.  Moreover, it is an experimental observation, repeated on many kinds of lasers, that the phase of a laser executes a diffusive motion proportional to the spontaneous emission rate.  Meanwhile, spontaneous emission, by Einstein's thermodynamic A and B coefficient argument, is
a necessary accompaniment of stimulated emission, and thus of laser
amplification \citeSOM{EinsteinDPG1916}.  It would thus be difficult to exclude spontaneous
emission, the archetype of a stochastic physical process, from the
description of laser phase diffusion.  

To describe the role of randomness in a Bell test, we first note that any given experiment can only test limited classes of HVTs.  Fully deterministic HVTs, which Bell named ``superdeterminism,'' cannot be tested by any Bell test, because in superdeterminism it is impossible to make free choices for the settings \cite{KoflerARX2014b} \citeSOM{LarssonJPA2014}.  In contrast, {stochastic} HVTs, i.e., those in which some events are unpredictable even in principle, can be tested.   Within this class, different HVTs will hold that some subset of: spontaneous emission, chaotic evolutions, thermal fluctuations, human decision-making, and any number of other arguably unpredictable processes, are stochastic.  A Bell test can exclude some of these classes of stochastic HVTs. 
%The RNGs described here, because they are based on laser phase diffusion, can test HVTs  in which phase diffusion is stochastic.  

{Directly observable} and thus theory-independent contributors to laser phase diffusion include spontaneous emission, Johnson noise in the injection current, and  carrier density fluctuation due to spontaneous carrier recombination. Because the phase is cyclic, if any one of these is sufficient to randomize the phase, it remains a pure random variable even when the others are known in advance.  A test using these generators can thus exclude HVTs in which any of these processes is stochastic.  In addition, a classical HVT in which the laser action is produced by classical electrons moving inside a the laser gain material would almost certainly be chaotic, allowing the test of another class of stochastic HVTs.

}

\section{Detection model}
When detected, $p_I(t)$ gives rise to an analog voltage signal $v(t) = [h * p_I](t) + \vD(t)$, where $h(t)$ is the impulse response of the detection system, $*$ indicates convolution, and $\vD$ is detector noise. Because the photoreceiver has a bandwidth much larger than the pulse repetition frequency, the signal mostly represents optical energy received within the last $\tauPD \sim \SI{100}{\ps}$, and thus from the present pulse.  Nevertheless, we need to take account of ``hangover error,'' i.e., delayed signal from previous pulses. 
% For example, comparing the baselines of Figs. \ref{fig:PulseShapes} and \ref{fig:InterferencePulses} we see that the random-amplitude pulses give rise to a noisier baseline.  This ``hangover error'' $\vhang$ provides a small but non-negligible offset to the true signal. 
We divide the impulse response as $h(t) = \hnow(t) + \hprior(t)$, where $\hnow(t)$ describes the short-time response of the detection system and is nonzero only for $0 \le t \le$ \SI{5}{ns}, i.e. for one pulse repetition period, whereas  $\hprior(t)$  is nonzero only for $t >$ \SI{5}{ns} and describes the delayed response.  We define $\vS \equiv \hnow * \pS$, $\vL \equiv \hnow * \pL$, $\vhang \equiv \hprior * p_I$ and the visibility $\vis$ through $\vphi \equiv \vis \cos \phi \sqrt{\vS \vL} \equiv \hnow * ( \cos \phi  \sqrt{ \pS \pL })$.  We then have
\bea
v &=& \hnow * p_I + \hprior * p_I + \vD \nne
%& = & \hnow * \pS +   \hnow * \pL +  \hnow * 2 \sqrt{\pS \pL} \cos \phi +  \hprior * p_I  \nnp  \vD \\ 
 % & = &
   \vS + \vL + 2 \vis \sqrt{\vS \vL} \cos \phi +   \hprior * p_I + \vD  \nne
 % & = & 
  \vS + \vL + \vphi + \vhang + \vD.
\eea
{Including the reference level noise $\vref$, the total untrusted noise is $\vc \equiv \vS +\vL + \vD + \vhang - \vref $. }

{
%\subsection{Completeness of the model}

It is natural to ask whether all relevant noises are included in $\vc$.  We note that, with the exception of $\vS$ and $\vL$, the individual noise contributions are each {defined to be} the total noise arising in some part of the conversion from optical to analog to digital.  The noise in converting from optical to analog is $\vhang+ \vD$.  $v_H$ is defined as all the fluctuation that comes from prior pulses due to slow response of the detection system, whereas $v_D$ is defined as all other classical noise arising in the detection process. The noise in converting from analog to digital is $\vref - \langle \vref \rangle$.  Again, this is defined to be all noise of that kind.  

$\vS - \langle \vS \rangle$ and $\vL - \langle \vL \rangle$ describe fluctuations in the individual pulse strengths as they leave the laser.  The fact that there are only two such sources comes from the fact that we are using an interferometer with two paths (short and long).  This describes the topology of the interferometer, and is unambiguous. 
}

\subsection{Distributions of  analog and digital signals}

The digitized signal $v = \vphi + \vc$ is the sum of $\vphi = \vrange \cos \phi$, where $\phi$ is trusted to be fully random and independent, and $\vc$, which is untrusted.  %We will refer to $v$ as a {\it corrupted independent source}.  
$\vphi$ is described by an arcsine distribution, with probability density function (PDF)
\be
\label{eq:PhiPDF}
P(\vphi = x) = 
\frac{1}{\pi \sqrt{\vrange - x} \sqrt{x + \vrange }} 
\ee
{if $|x| < \vrange$ and zero otherwise.} The cumulative distribution function is
%\be
%\label{eq:PhiPDF}
%P(\vphi = x) = \left\{  
%\begin{array}{cl}
%\frac{1}{\pi \sqrt{\vrange - x} \sqrt{x + \vrange }} & |x| < \vrange \\
%0 & {\rm otherwise}
%\end{array}
%\right.
%\ee
\be
P(\vphi \le x) \equiv \int_{-\infty}^{x} dx\,P(x) = \frac{2}{\pi} \arcsin \sqrt{\frac{1}{2} + \frac{x}{2\vrange}},
\ee
{for $|x| \le \vrange$.}
%\bea
%P(\vphi \le x) &\equiv& \int_{-\infty}^{x} dx\,P(x) \nne 
%\left\{  
%\begin{array}{cl}
%1 & x \ge \vrange\\
%\frac{2}{\pi} \arcsin \sqrt{\frac{1}{2} + \frac{x}{2\vrange}} & |x| < \vrange \\
%0 &  x \le -\vrange 
%\end{array}
%\right. .
%\eea
We assume $\vc$ is available to the distant particle and ask what distribution the particle would predict for $v$, knowing $\vc$.  This describes a conditional distribution $P(v = x| \vc)$ equal to $P(\vphi = x -\vc)$, which is simply $P(\vphi=x)$  shifted by $\vc$.  When digitized as $d= \theta[ \vphi + \vc]$, where $\theta$ is the Heaviside function,  the conditional distribution for $d$, i.e. the probability mass function, is
\bea
P(d=0) &=& P(\vphi + \vc \le 0) = P(\vphi \le - \vc) 
\nne \frac{2}{\pi} \arcsin \sqrt{\frac{1}{2} - \frac{\vc}{2\vrange}},
\eea
{for $|\vc| \le \vrange$.} $P(d=1)$ is then $1-P(d=0) =  \frac{2}{\pi} \arcsin \sqrt{\frac{1}{2} + \frac{\vc}{2\vrange}}$ as given in Eq. (3). %\ref{eq:PredFromVc}).  

\section{Randomness extraction %from corrupted independent sources.
}

Consider two partially-random bits $a = \theta[\vphi^{(a)} + \vc^{(a)}]$ and $b = \theta[\vphi^{(b)} + \vc^{(b)}]$ generated as above, where superscripts $^{(a)}$ and $^{(b)}$ indicate distinct realizations of the variables.  Because $\vc^{(a)}$, $\vc^{(b)}$ are not trusted to be independent, the joint probability  $P(a=x,b=y)$ is not in general separable.  Nevertheless, the conditional joint probability separates: $P(a=x,b=y | \vc^{(a)}, \vc^{(b)}) = P(a=x|\vc^{(a)})P(b=y|\vc^{(b)})$, because the conditioned $a$ and $b$ are a function only of the independent $\vphi^{(a)}$ and $\vphi^{(b)}$, respectively.  This allows us to use the properties of independent random variables in computing predictability bounds.

For a partially-random bit $d$, the predictability is $\Pred(d) \equiv \max[ P(d=0), P(d=1)]$.
%, where $P(d=x)$ is the probability that $d$ takes on the value $x$. 
 An ideal random bit has predictability $1/2$, and writing $\Pred(d) = \frac{1}{2}(1+\epsilon_d)$ we find the error $\epsilon_d = 2 \Pred(d) - 1$.  Considering two partially-random bits $a$ and $b$, with predictabilities $\Pred(a), \Pred(b)$, respectively, $\Pred(a \oplus b)$, the predictability of $a \oplus b$ 
 %, understood to be conditioned on the untrusted observables, 
 is the larger of 
 \bea
 P(a \oplus b = 0) &=& P(a=0) P(b=0) + P(a=1) P(b=1) \nonumber 
 \eea
 and
 \bea
 P(a \oplus b = 1) &=& P(a=0)P(b=1) + P(a=1)P(b=0), \nonumber
 \eea from which we find
\be
\Pred(a\oplus b) = \Pred(a) \Pred(b) + [1-\Pred(a) ] [1-\Pred(b)]
\ee  
or $2\Pred({a\oplus b}) -1 = [2 \Pred(a)-1][2 \Pred(b) -1]$, so that 
\be
\epsilon_{a \oplus b} = \epsilon_a \epsilon_b.
\ee
Moreover, if  $\epsilon_a \le\epsilon_{a,\rm max} $ and $\epsilon_b \le \epsilon_{b,\rm max} $
%$\Pred(x) \ge \Pred\supmax(x)$ and $\Pred(b) \ge \Pred\supmax(b)$,  
i.e. if the errors are bounded from above, then 
%because $\Pred(a\oplus b)$ is monotonically increasing in both $\Pred(a)$ and $\Pred(b)$, we have an upper bound $2 \Pred(a \oplus b) -1 \le 2\Pred\supmax(a\oplus b) -1 \equiv (2\Pred\supmax(a)-1) (2\Pred\supmax(b)-1)$.  More conveniently,
\be
\label{eq:epsilonaplusb}
\epsilon_{a \oplus b} \le %\epsilon\supmax_{a \oplus b} \equiv 
\epsilon_{a,\rm max} \epsilon_{b,\rm max}
\ee
because 
%$\Pred(a\oplus b)$ and thus 
$\epsilon_{a \oplus b}$ is monotonically increasing in both  $\epsilon_a$ and $\epsilon_b$. When computing $x_{i+k} \equiv x_{i} \oplus d_{i+1} \oplus \ldots \oplus d_{i+k}$, where $\epsilon_{d_j} \le \epsilon_{d_j,\rm max}$, repeated application of Eq. (\ref{eq:epsilonaplusb}) gives  
\be
\epsilon_{x_{i+k}} \le \Pi_{j=1}^k \epsilon_{d_{i+j},\rm max}.
\ee
This shows  an exponential approach to ideal randomness. 
Note that $x_{i}$, which is due to events in the past light-cone of the detection, as shown in Fig.~1,
%~\ref{Fig:setupMain}~a)), 
contributes no randomness and $\epsilon_{x_{i}} = 1$. 

\newcommand{\tsc}{t_s^{(c)}}
\newcommand{\tss}{t_s^{(s)}}

\newcommand{\vOhg}{\vO}
\newcommand{\vOmg}{\vO'}
\newcommand{\vOlg}{\vO''}

 \begin{table}[!hbp]
\caption{\label{tab:Noises}Measured noise statistics.  All voltages in \SI{}{mV}. $\vOhg$, $\vOmg$ and $\vOlg$ indicate the oscilloscope noise at gains of \SI{50}{mV/division}, {\SI{100}{mV/division}} and  \SI{200}{mV/division}, respectively.  $\vL + \vhang+ \vD +\vOhg$ is measured using an interrupted pulse train, as in Section \ref{Sec:Hangover}.  $\vref + \vOhg$ is measured using the x-y method of Sec. \ref{Sec:XY}.  $\Delta \vphi$ is determined from the fit of Fig.~3. %\ref{Fig:Validation}~a). 
 }
\centering
\begin{tabular*}{1\linewidth}{R{0.4\linewidth} R{0.1\linewidth} R{0.01\linewidth} | R{0.15\linewidth} R{0.1\linewidth} R{0.1\linewidth}}
\toprule
measured variables &  r.m.s. dev. & &  derived variables & mean & r.m.s. dev.   \\
\hline
$\vOhg$ & 1.7 & \\
$\vD+\vOhg$ & 1.9 & & $\vD$ & & 0.8 \\
$\vS + \vD + \vOhg$ & 2.2& & $\vS$ & 251 & 1.1 \\
$\vL + \vD + \vOhg$ & 2.3& & $\vL$ &  251 & 1.3  \\
$\vL + \vhang+ \vD +\vOhg$&  4.4 &  & $\vhang$    & &  3.8  \\
$\vOmg$ & 3.4 & \\
$\vref + \vOmg$ & 8.4& & $\vref$ & & 7.7  \\
$\vOlg$ & 7 & \\
$\vphi + \vS + \vL +  \vD + \vhang + \vOlg$ & 354 & & $\Delta \vphi$ & 483 &    \\
\hline
%derived noise & mean & dev  & min & max & N \\
\toprule
 \end{tabular*}
 \label{Tab:NoiseStatistics}
 \end{table}

\begin{figure*}[t]
\centering
\includegraphics[width=0.85 \linewidth]{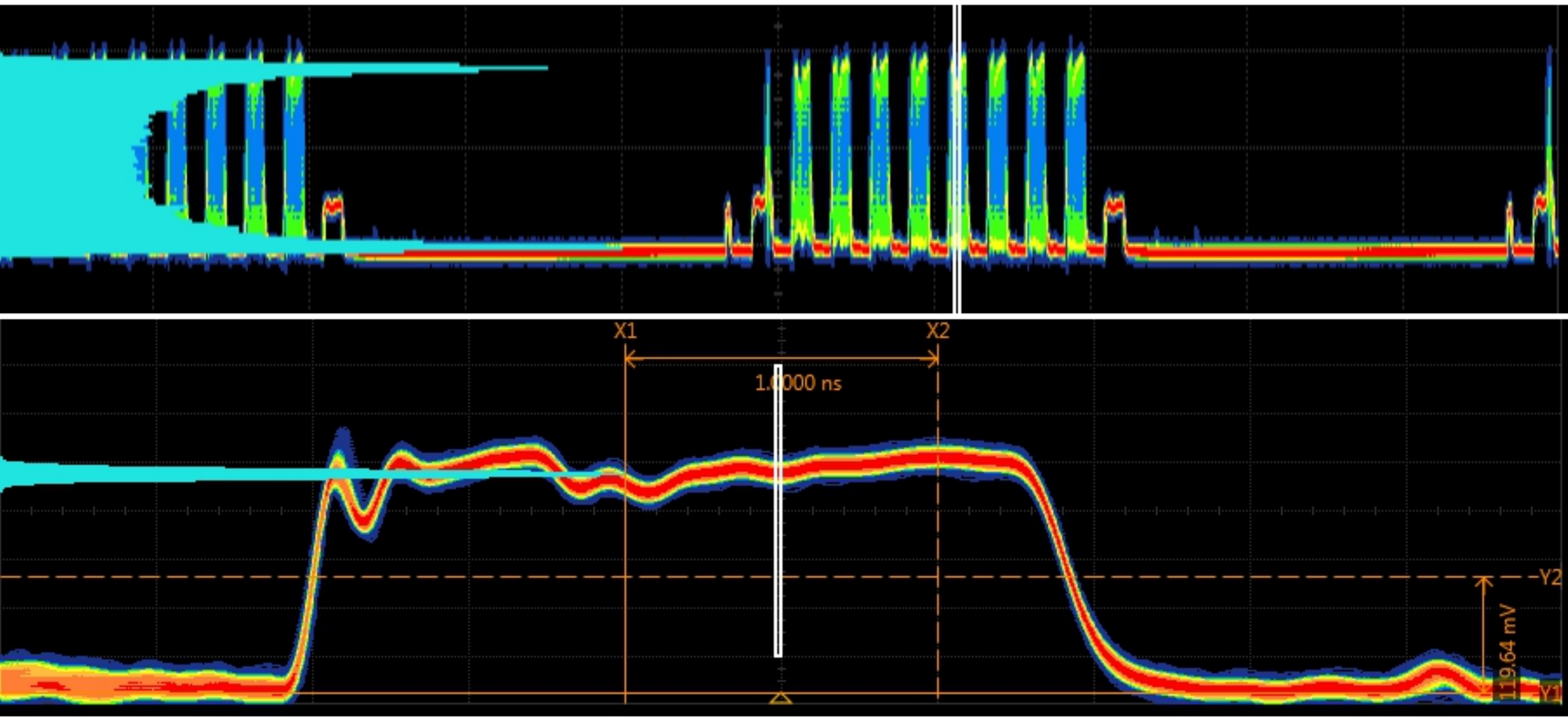}
\caption{Measurement of ``hangover error.''  Upper image shows persistence mode oscilloscope trace  of trains of 9 optical pulses giving rise to 8 strongly interfering pulses followed by one pulse without interference, shown in higher resolution in the lower image.  Pulse repetition period is 5 ns.  Teal histograms on left vertical axis show data sampled in the region defined by the white cursors.  ``Hangover,'' i.e., remaining variation from previous pulses, is visible at the start of the lower trace, and decreases approaching the sampled point.  } 
\label{Fig:Hangover} 
\end{figure*}

\section{Noise measurements} 
Statistics are measured with an AC-coupled oscilloscope  (Agilent Infinitum MSO9404A)  with 4 GHz input bandwidth and 8 bit resolution. Statistics are acquired as histograms of sampled voltages within a {50 ps} window, as shown in {Fig.~2~b) \& {c)}.}  All signals in Table \ref{tab:Noises} except $\vO$ and $\vref + \vO$ were measured at the photodiode output. To measure $\vL + \vD + \vO$ or $\vS + \vD + \vO$ we block the short or long path of the interferometer, respectively.  To measure hangover errors, we use a fast analog switch (Mini-circuits {ZASWA-2-50DR+)} to periodically block the train of current pulses to the LD, see Section \ref{Sec:Hangover}.  Immediately following the turn-off, a single long-path pulse arrives to the photodiode with no corresponding short-path pulse, thus producing a voltage $\vL + \vhang + \vD + \vO$.  The full interferometer produces $\vphi + \vS + \vL + \vD + \vhang + \vO$.  

The reference voltage $\vref$ is studied {in} two ways.   1) We measure the width of the comparator transition threshold, by correlating the digital output to the random analog input (see subsection \ref{Sec:XY}).  2)  direct measurement of $\vref$ with the oscilloscope.  These two methods give consistent results for $\sigma_{\vref}$. In contrast to $\vphi$, all technical noises contributing to $v$ appear normally distributed with histograms obeying the ``68-95-99.7'' rule.  

We simulated Eq. (2) by generating normally distributed random numbers with means and r.m.s. deviations given by the measurements shown in Table \ref{tab:Noises}. The phase was simulated to be fully randomised. The resulting distribution was fitted to an arcsine distribution, as shown in Eq. (\ref{eq:PhiPDF}), finding $\langle 2\vrange \rangle =  \SI{966}{mV}$. The fit is shown in Fig.~3. In calculating the $6\sigma$ bound on $\Pred$, we adjust this value downward by a multiplicative factor $\sqrt{1 - 6\sigma_{\vS}/\vSbar}\sqrt{1 - 6\sigma_{\vL}/\vLbar} \approx 0.971$ to conservatively account for fluctuations in $\vrange$.

\begin{figure*}
\centering
\includegraphics[width=0.45 \linewidth]{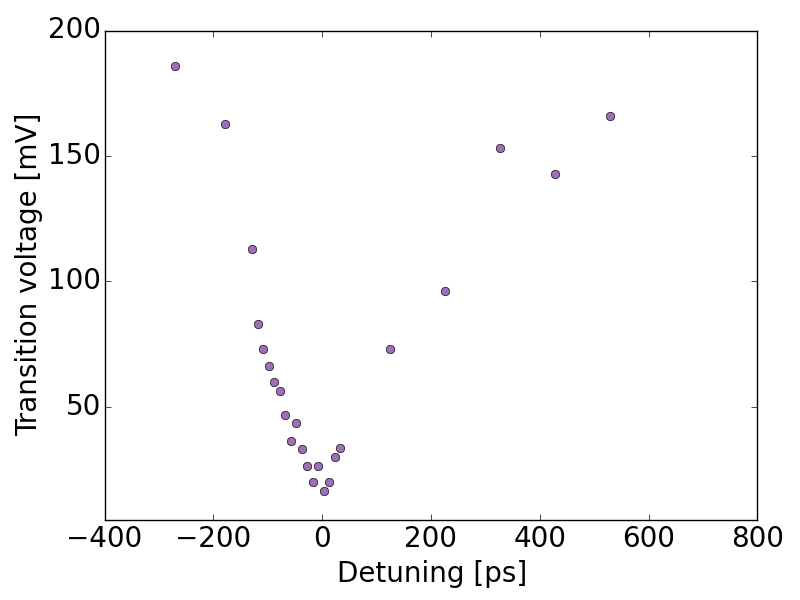}
\includegraphics[width=0.45 \linewidth]{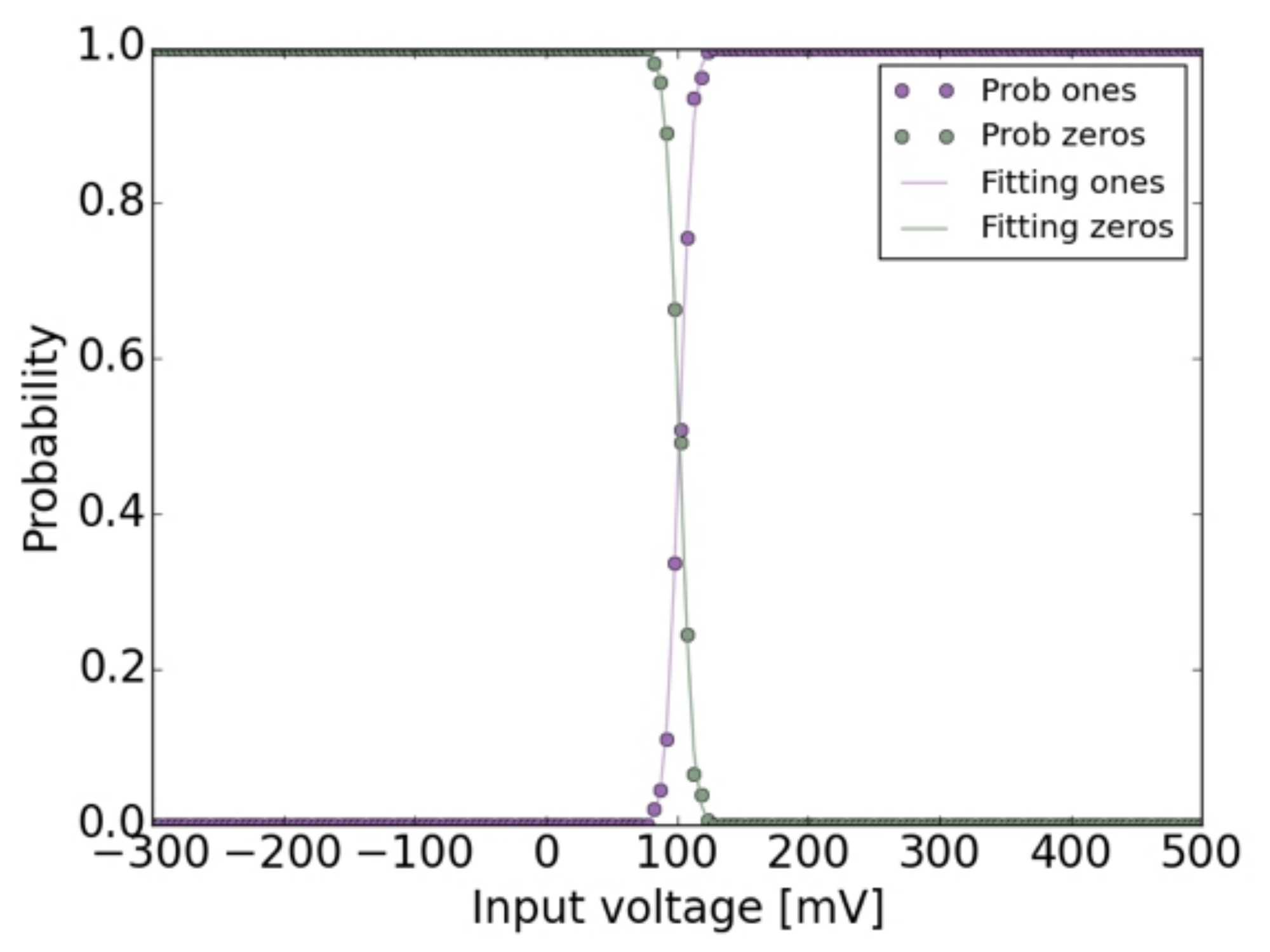}
\caption{(Left) Probability of getting an output zero or one for each input analog voltage $v_2$. As observed, there is a small region near the threshold region in which input values can give zero or one with certain probabilities. The transition probability from $0\rightarrow 1$ ($1\rightarrow 0$) is in excellent agreement with the integral of a Gaussian with $\sigma_{\vref} \le$ \SI{7.7}{mV}. (Right) Measured noise as a function of the detuning between the oscilloscope sampling time and the flip-flop sampling time. We emphasise this measurement is conservative, as the measured error is necessarily larger than the real error.\label{Fig:xycomp} }
\end{figure*}

We find $\langle \vc \rangle$,  the mean of the combined noises,  from 
%$\langle \vc \rangle \approx [P(x=1) - 1/2] \pi \vrange/2 = \mtext{??}$ 
$\langle \vc \rangle = 2\vrange[\langle \sin^2 (P_1 \pi/2) \rangle - 1/2] 
\approx 2\vrange[ \sin^2 (\langle P_1 \rangle \pi/2) - 1/2] = \SI{3.0}{mV}$, using the measured $\langle P_1 \rangle = 0.50035$.  The approximation of $\langle \sin^2 (P_1 \pi/2)\rangle$ is justified in light of Eq. (3) %\ref{eq:PredFromVc}) 
and the observed $|\vc|  \ll \vrange$.  

\subsection{Measurement of hangover errors}
\label{Sec:Hangover}
To measure the hangover errors, we periodically interrupt the modulation of the {LD} using an RF switch (Mini-Circuits ZASWA-2-50DR+) at 10 MHz.  This generates a train of optical pulses at the output of the laser.  Due to the relative path difference in the interferometer, three different types of pulses emerge: (i) the first output pulse, which contains only a short-path contribution and experiences no interference, (ii) the intermediate pulses, which contain both short and long-path contributions and show interference and (iii) the last pulse, which contains only a long-path contribution and thus shows no interference.  This last pulse also contains any delayed response, i.e. ``hangover,'' from previous pulses.  By measuring the statistical behavior of this last pulse, and comparing against the long-path signal obtained by blocking the short-path, we can recover the contribution from hangover errors. This is illustrated in Fig. \ref{Fig:Hangover}, which shows a train of nine optical pulses.  For the data reported in Table \ref{Tab:NoiseStatistics} trains of ten pulses were used.

%\section{Physical characterisation of the device}
\subsection{Noise in the reference level: Input-Output analysis}
 \newcommand{\Tup}{\begin{cal} T \end{cal}}
  \newcommand{\Int}{\begin{cal} I \end{cal}}
\label{Sec:XY}
In addition to a direct measurement of $\vref$, the comparator reference voltage, we also measured the input-output relation, which quantifies the performance of the {1-bit} analog-to-digital conversion (ADC). Note that while an ideal comparator converts each analog input into an unambiguous digital output, a real comparator has noise and therefore the conversion of input values near the reference voltage might contain some uncertainty. In order to  quantify this ``transition voltage'' range, it is not sufficient to measure the reference noise, as any extra effect occurring inside the comparator itself or lack of knowledge of the performance of the device would be neglected. We emphasise this measurement makes no assumptions at all about the circuit, making the measurement outcome ultimately transparent. As shown in Fig. \ref{Fig:comp_setup}, the measurement setup is as follows:  we use a 3 dB splitter (Mini-Circuits ZFRSC-183-S+) after the photodetector to get two copies of the output analog random amplitude $v_{1,2}$. We send $v_1$ to the comparator input and $v_2$ to the oscilloscope, which also records the output of the first flip-flop, i.e., the comparator output latched exactly at the same point the oscilloscope is sampling. In order to match the sampling points of the oscilloscope and the flip-flop, we sweep the sampling point of the oscilloscope in steps of 10 ps until we find the best agreement, see Fig.~\ref{Fig:xycomp}~(left). Note that having the oscilloscope sampling at exactly the same point as the flip-flop is critical for this measurement. The analysis shows a small but finite overlap between range of input values giving low versus high output values near the threshold voltage. Computing the frequency of high and low outputs for each input value $v_2$, we obtain the conversion probabilities $P(d=x|v_2)$ for $x\in \{{\rm 0,1 }\}$.

\begin{figure}[b]
\centering
\includegraphics[width=0.8\linewidth]{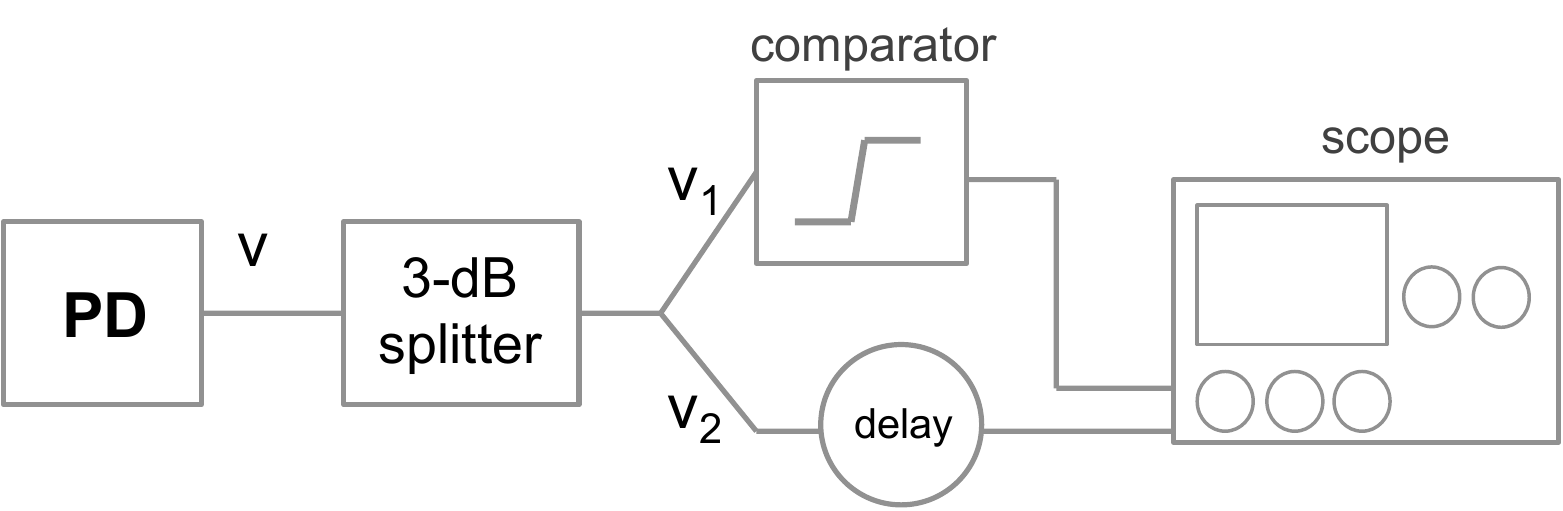}
\caption{\label{Fig:comp_setup}Schematic for the characterization of the comparator chip. The signal from the photodetector (PD) is split in two by means of a 3-dB splitter. One of the signals is sent to the comparator and the other to the scope, and delayed using the skew function of the scope. The sampling point of the comparator can be tuned with $10$ ps resolution. By capturing the digital signal output of the comparator and the analog signal recorded by the scope, we can measure the performance of the comparator, e.g. the probability of the output conditioned on the input.}
\end{figure}

The measurement suffers from some limitations. In an ideal scenario, we would need (i) the two outputs of the 3-dB splitter to be identical and (ii) the sampling point of the oscilloscope and flip-flop too. In practice, unfortunately, (i) the two outputs of the 3-dB splitter are not identical but their difference follows a Gaussian distribution with 0 mean and 1.32 mV rms deviation, and (ii) the timing precision is limited to 10 ps. Note that the presence of both limitations are conservative from the measurement point of view, i.e., the real error will {always be} smaller than the measured error. The narrowest observed transition is depicted in Fig.~\ref{Fig:xycomp}~(right), and shows an r.m.s. width of \SI{8.4}{mV}. Considering that the oscilloscope noise is 3.4 mV rms, we can place an upper limit: $\sigma_{\vref} \le$ \SI{7.7}{mV}.% which is considerably above the direct measurement result of \mtext{$\sigma_{\vref} \le$ \SI{7.7}{mV}}.

\section{Upper and lower bounds on the freshness time}

As illustrated in Fig. 1 of the main article, the window for randomness generation is bracketed on the early side by the requirement for space-like separation from the distant detection, and on the late side by the requirement for space-like separation from the pair generation.  Ensuring the random events fall in this window requires both upper and lower bounds on the freshness time.  We measure timing of relevant events using a differential probe and a 20 GSa/s real time oscilloscope (Agilent Infinitum MSO9404A). As shown in Fig. \ref{Fig:timing_diagram}, we measure three  delays in the circuit: (i) $t_{1}$: from the modulation of the laser, measured directly on the pins of the laser, to the output of the photodetector, (ii) $t_{2}$: from the output of the photodetector to the input of the XOR gate, and (iii) $t_{3}$: from the input of the XOR gate to the CML output connector. We split in three intervals for traceability of the signal while travelling through the electronics.

\begin{figure}[b]
\centering
\includegraphics[width=1\linewidth]{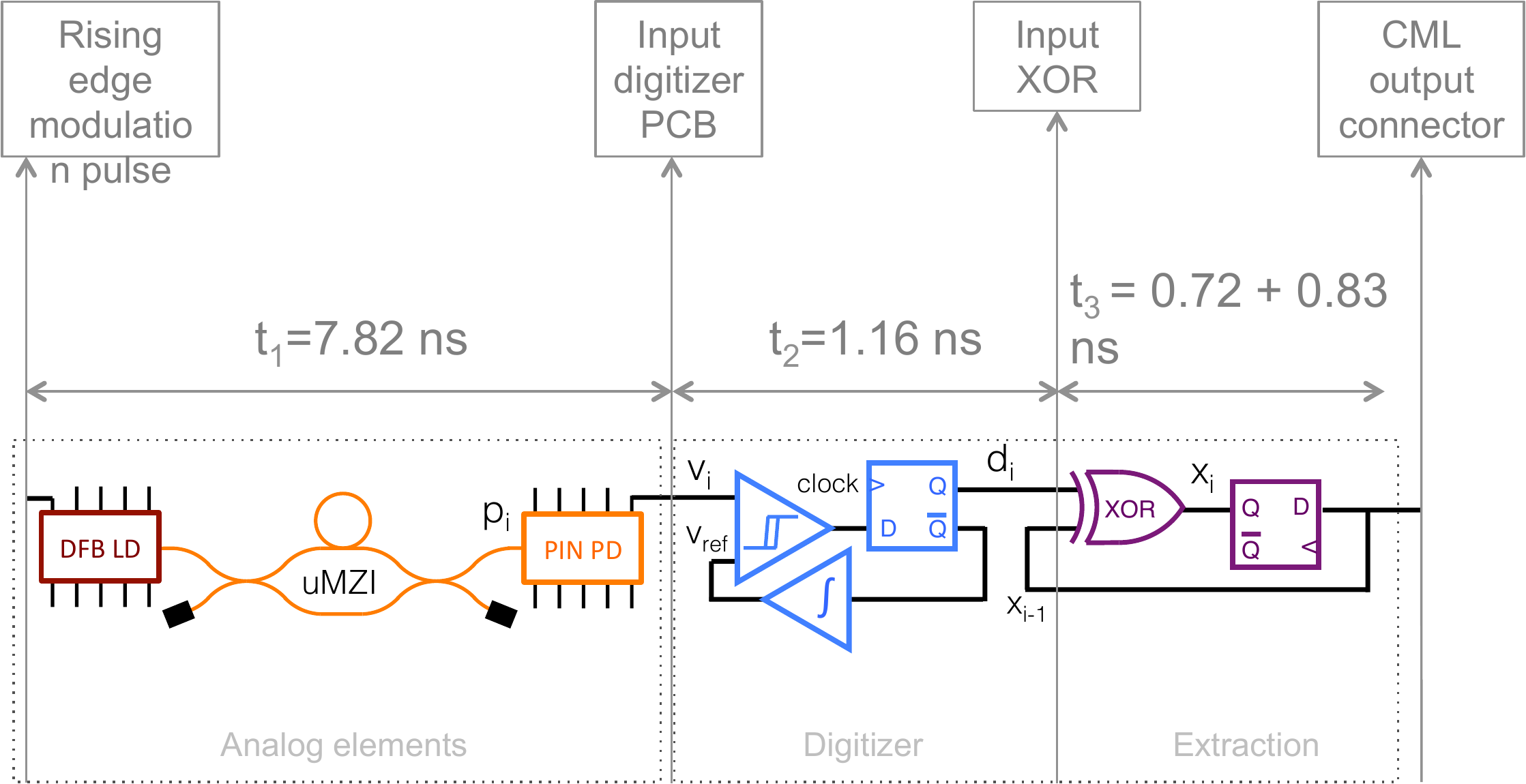}
\caption{\label{Fig:timing_diagram}(color online) Schematic of the random number generator with timing information as directly measured by a 20 GSa/s real time oscilloscope. $t_{1}$ describes the time from rising edge of the current pulse until rising edge of the PIN PD output pulse, $t_{2}$ describes the time from the PIN PD pulse to the input of the XOR gate, and $t_{3}$ the time from the input of the XOR gate to the output SMA port. }
\end{figure}

\begin{figure}[t]
\centering
\includegraphics[width=0.7\linewidth]{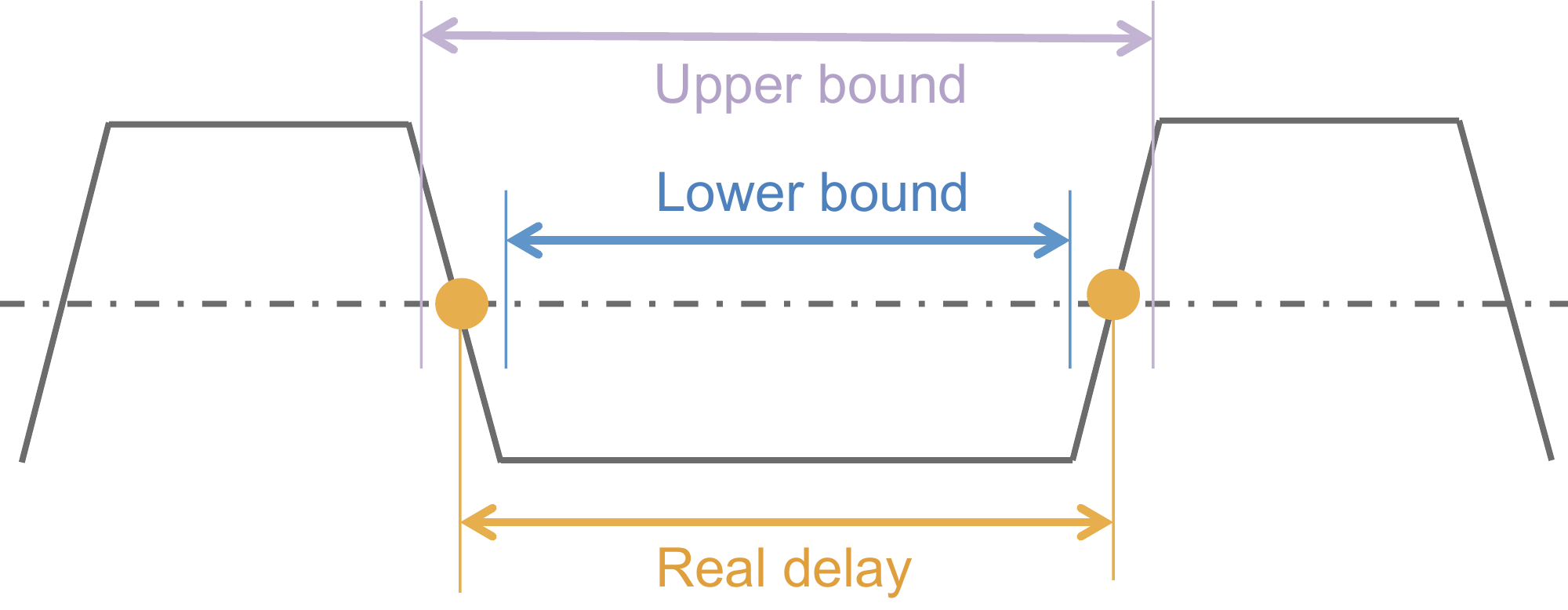}
\caption{\label{Fig:cursors}(color online) For a given captured trace, the best-guess estimate for the zero-crossing time is found by interpolation between the 50 ps samples of the 20 Gsps oscilloscope, with systematic uncertainty (orange circles). To obtain the upper bound (purple) and lower bound (orange) for the zero-crossing time, we add or subtract the half-width of the transition.  Statistical uncertainty is accounted separately from the statistics of $10^7$ traces collected in the oscilloscope's persistence mode. }
\end{figure}

 \begin{figure}[!hb]
\centering
\includegraphics[width=0.75\linewidth]{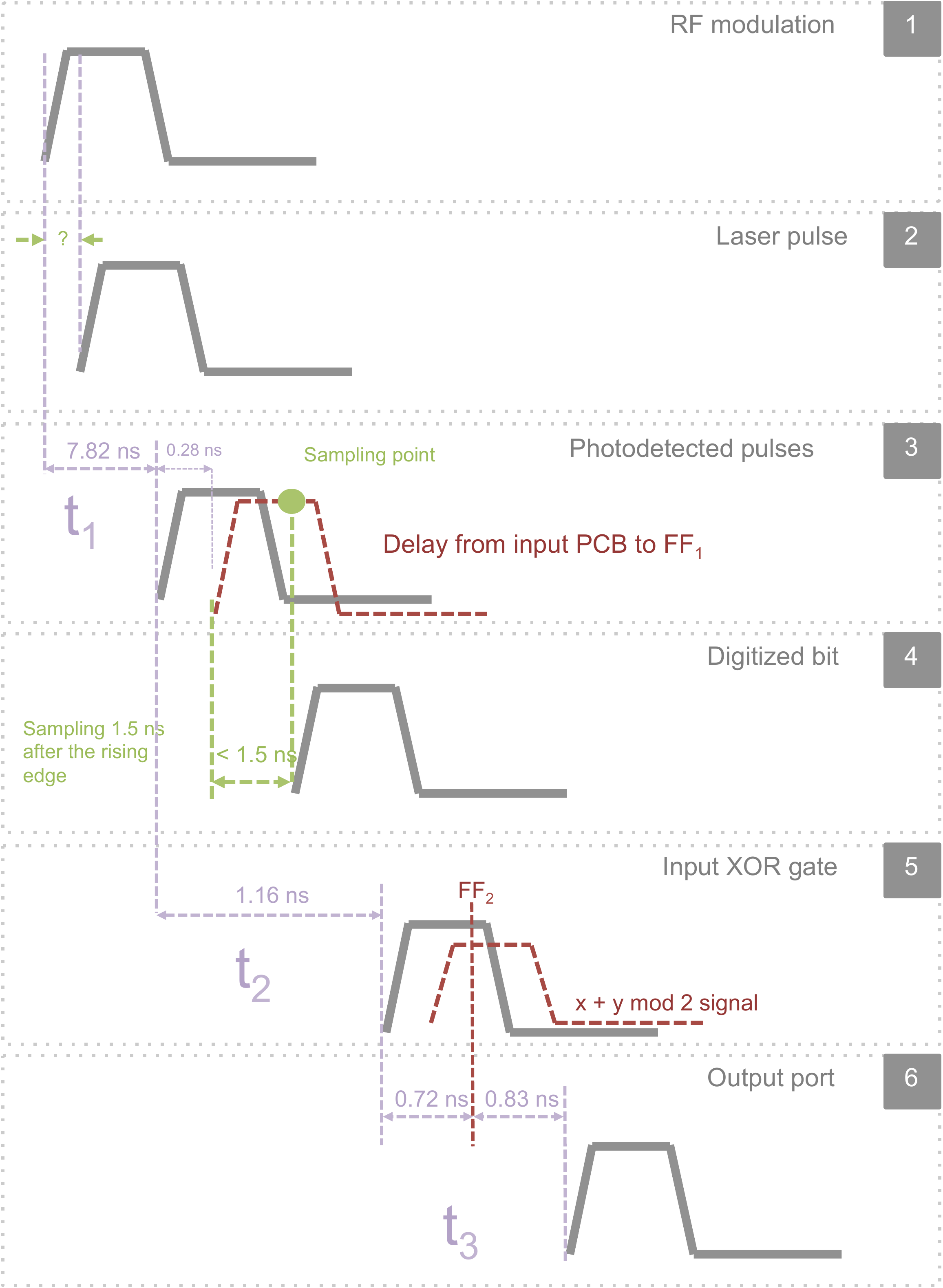}
\caption{\label{Fig:timing_measurements}(color online) Details of the sequence of measurements to obtain the freshness time. We start by measuring the modulation pulse directly at the laser diode pins (1). The LD generates an optical pulse some time after the modulation pulse is applied (2). Then, the signal is photodetected and measured (3). The difference between (1) and (3) gives $t_{1}$. Then, the electrical pulse is digitised by a fast comparator and latched by a first flip-flop (4), and then propagates until the input of an XOR gate (5). At the input of the XOR gate we can tap the signal and measure the arrival time. The difference between (3) and (5) gives $t_{2}$. Finally, the signal goes into the randomness extraction circuitry (see main text for more details), and propagates to the output, where we measure the arrival time again (6). The difference between (5) and (6) gives $t_{3}$. }
\end{figure}

All delays are quantified by capturing traces and using cursors to identify the zero-crossing times of relevant edges.  As shown in Fig. \ref{Fig:timing_diagram}, for each transition we identify a best guess $t_{i}^{({\rm best})}$ limited by the uncertainty of the interpolation between $50$ ps samples.  We find $t_{1}^{(\rm best)}=7.82$ ns, $t_{2}^{(\rm best)}=1.16$, and $t_{3}^{(\rm best)}=1.55$ ns.  

Measuring intervals on the oscilloscope can give a systematic error, which we now bound.  In light of the $\sim 100-150$ ps 10\%-90\% rise and fall times of the transitions, it is extremely improbable that we misjudge the location of the edge by 100 ps or more, as it would mean placing the best guess outside of the transition region.  As illustrated in Fig. \ref{Fig:cursors}, we calculate upper and lower bounds for the transition times as $t_{i}^{({\rm ub})} = t_{i}^{({\rm best})} + 100$ ps and $t_{i}^{({\rm lb})} = t_{i}^{({\rm best})} - 100$ ps, respectively.
%, being $100$ ps already an extremely conservative number for systematic errors due to interpolation.
%For finding an upper-bound $\tau_{\rm f}^{(ub)}$, we place the cursors further away from the $50$\% level of the signal, and viceversa for finding a lower-bound $\tau_{\rm f}^{(lb)}$. 
Combining the three measured intervals we find $\tau_{\rm f}^{({\rm ub, sys)}} = \sum_{i=1}^3 t_{i}^{({\rm ub})} = 10.87$ ns and $\tau_{\rm f}^{(\rm lb,sys)} = \sum_{i=1}^3 t_{i}^{({\rm lb})} =10.21$ ns.   Fig. \ref{Fig:timing_measurements} shows the sequence of measurements performed.

The jitter of the signal plus the jitter of the oscilloscope is quantified by accumulating $10^7$ traces using the persistence mode of the oscilloscope.  Measuring the rising edge of the signal at the photodiode output (as in Fig. \ref{Fig:Hangover}, e.g.), we observe that all of the traces fall within a $125$ ps temporal window; there is no recorded value outside of this window.  Making the hypothesis that events outside of this window occur with probability at least $1.4 \times 10^{-6}$, we expect $\ge 14$ events on average outside of this window.  A Poisson distribution with this mean predicts our observed zero events with probability $8 \times 10^{-7}$.  We can thus reject the hypothesis and assign a $p<1.4 \times 10^{-6}$ confidence to the $125$ ps window for the measured zero-crossings.  We note that this does not count the contribution of the oscilloscope to the jitter and is thus conservative.  
%We note that this measurement, in contrast to the single-trace timing used to find the best guess, makes use of the triggering of the oscilloscope to align the many traces.  Due to the 50 ps sampling period and asynchronous operation of the oscilloscope, this systematically broadens the distribution of observed zero-crossing times by at least the 50 ps uncertainty of the triggering.  We can thus assign the same $p<1.4 \times 10^{-6}$ confidence to a window of $\le 75$ ps width for the zero-crossings of the signal itself.    
The freshness time combines intervals from three cascaded measurements.  Adding three such windows, the conservatively estimated window for the full process is $3\times 125 = 375$ ps.  Half of this, 187.5 ps,  can be assigned to the upper limit, and half to the lower.  To have a round number, we define the jitter bound to be $\tau_{\rm jitt} = 200$ ps. 

The lower and upper bounds for the freshness time of a single bit, including statistical and systematic errors, conservatively estimated, are then
\bea
\tau_{\rm f}^{(\rm lb,sys)} - \tau_{\rm jitt}  &< \tau_{\rm f} <& \tau_{\rm f}^{(\rm ub,sys)} + \tau_{\rm jitt} \\ 
%\tau_{\rm f}^{(\rm lb)} &< \tau_{\rm f} <& \tau_{\rm f}^{(\rm  ub)}\\
10.01 \text{ ns} &< \tau_{\rm f} <& 11.07 \text{ ns}
\eea

We also observe that the rms width of the jitter is $\sigma_{\rm jitt} < 40$ ps.  If the jitter can be considered normally distributed, $\tau_{\rm jitt}$ is then at least $5 \sigma_{\rm jitt}$, and could be made considerably stronger, e.g. at least $10 \sigma_{\rm jitt}$, with only a 200 ps broadening of the bounds given above. 

The operating frequency of the quantum random number generator is nominally $200$ MHz, which corresponds to a cycle time of $5$ ns.  The clock of the RNG can be derived from an external reference via a phase-locked loop, or internally generated from a quartz oscillator (Analog Devices AD9522/PCBZ), in either case introducing a timing uncertainty that is nominally $<1$ ps and  negligible on the scale of the other uncertainties.    The freshness time for $k$ events is therefore given by
\bea
\tau_{\rm f}^{(k)} = \tau_{\rm f} + 5 \times (k-1) \text{ ns}.
\eea

\section{Statistical testing}
A fast digitizer (Acquiris U1084A) is used to acquire the RNG output for statistical testing. Because of memory limitations, data is acquired in runs of $\approx\SI{8}{Mb}$ each and concatenated. 

\subsection{Sample autocorrelation}

The two-point autocorrelation is $\Gamma_\corrvar(\corrdist) \equiv \langle \corrvar_i \corrvar_{i+\corrdist} \rangle - \langle \corrvar \rangle^2$, where $\corrdist$ is the correlation distance.  
%$\Gamma_\corrvar$ is a convenient experimental measure used in some analyses of Bell tests \citeSOM{Kofler:2014tn}.  
To obtain experimental autocorrelations from a sample of $N$ bits as in Fig.~5,%\ref{Fig:Validation}~b),
we compute the unbiased estimator for $\Gamma_\corrvar(\corrdist)$ 
\begin{equation}
\hat{\Gamma}_\corrvar(\corrdist) \equiv \frac{1}{N-\corrdist}\sum_{i=0}^{N-\corrdist} \corrvar_i \corrvar_{i+\corrdist} - \left[\frac{1}{N}\sum_{i=0}^{N} \corrvar_i \right]^2.
\end{equation} 
For a perfect coin and large $N$, the estimator has rms statistical uncertainty $\sigma_{\hat{\Gamma}} = 1/4\sqrt{N}$.

Considering the output of the RE, with $x_i = x_{i-1} \oplus d_i$, where $d_i$ are raw bits, we note that $x$ can be described as a symmetric two-state machine that changes state whenever $d = 1$.  This symmetry guarantees the long-time average $\langle x \rangle = 1/2$, except if $d$ is deterministic.  

\newcommand{\Probb}[1]{Pr[{#1}]}
\newcommand{\ppar}{\epsilon}

We estimate $\Gamma_\corrvar(\corrdist)$ when $d$ has bounded predictability, i.e. $\max [ P(d_i=0),  P(d_i=1)] \le \frac{1}{2}(1+\epsilon_{\rm max})$ for all $i$.  
We note that $\corrvar_i \corrvar_{i + \corrdist}$ is only nonzero for $\corrvar_i = \corrvar_{i + \corrdist} = 1$, so we can evaluate $\langle \corrvar_i \corrvar_{i + \corrdist} \rangle$ as the probability $P(\corrvar_i = 1)=\frac{1}{2}$ times the conditional probability $P(\corrvar_{i+\corrdist} =1 | \corrvar_i = 1)$.  This latter is the probability of an even number of $d=1$ raw bits between $i$ and $i+\corrdist$.  Subject to the bound, this is maximized when $P(d_i=0) = \frac{1}{2}(1+\epsilon_{\rm max})$ for all $i$.  Counting the possible ways to have an even number, where $n$ is the number of bits with $d=1$, we find
%Writing $P_1 \equiv P(d=1)$, we find  
\newcommand{\nodd}{n_{}}
\bea
\Gamma_\corrvar(\corrdist) & = & \langle \corrvar_i \corrvar_{i + \corrdist} \rangle - \langle \corrvar\rangle^2
\nn
& \le & \frac{1}{2^{k+1}}\sum_{\nodd=0}^{\lfloor \corrdist/2\rfloor} \binom{\corrdist}{2\nodd} (1 - \epsilon_{\rm max})^{2\nodd}  (1 + \epsilon_{\rm max})^{\corrdist-2\nodd} - \frac{1}{4}
\nne \frac{1}{4} \epsilon_{\rm max}^\corrdist.
%\nne \frac{1}{2}\sum_{i=0}^{\lfloor \corrdist/2\rfloor} \binom{\corrdist}{2i} P_1^{2i} (1-P_1)^{\corrdist-2i} - \frac{1}{4}
%\nne \frac{1}{4} (2P_0-1)^\corrdist
\eea

\subsection{Statistical test batteries}

Several statistical batteries are used to test the quality of the output, including the TestU01 Alphabit battery \cite{LEcuyerACM2007}, the NIST SP800-22  battery \cite{RukhinNIST2010tech}, and the Dieharder battery \cite{BrownWEB2004b}.  The results are consistent with ideal randomness for $k=3$ and above.  

Due to the high output rate of the RNG, testing was limited by computation speed for the various tests.  In this regard Alphabit has a significant advantage, as it was designed for testing physical RNGs, without the more computationally-intensive tests used for pseudo random number generators.  For example, testing a $1.5$ Gb sequence with the NIST battery takes more than 3 hours on a desktop computer whereas the Alphabit battery takes one minute.

\subsubsection{Dieharder tests}
We ran the dieharder test with default settings.  Results are shown in Table \ref{Table:dieharder}.

\newpage 
\begin{table*}[!htp]
\caption{\label{Table:dieharder}Summary of results for the entire Dieharder battery for k=3.   In column {\tt n tuple}, the notation $1\ldots x$ indicates that the test was repeated with the {\tt n tuple} setting covering this range.   As shown, the diehard parking lot test showed a {weak} value, i.e., an inconclusive result, in the initial run. For an ideal source, it is expected that $\sim 1$ weak value will appear in any given full test run of the suite.  As recommended, we re-run the weak test with the option {\tt -Y}, which increases {\tt p samples} until a clear result (passed or failed) emerges.  The test was passed.  }
\centering
\begin{tabular*}{0.9\linewidth}{L{0.2\linewidth} C{0.1\linewidth} C{0.1\linewidth} C{0.1\linewidth} C{0.2\linewidth} C{0.2\linewidth} }
\toprule
\textbf{test name} & \textbf{n tuple} & \textbf{t samples} & \textbf{p samples} &  \textbf{$p$-value} & \textbf{Assessment} \\
\hline
diehard birthdays&   0&       100&     100&0.86914871&  PASSED \\
diehard operm5 &   0&   1000000&     100&0.62710352&  PASSED\\ 
diehard rank 32x32&   0&     40000&    100&0.49138373&  PASSED\\ 
diehard rank 6x8&   0&    100000&     100&0.46907910&  PASSED\\ 
diehard bitstream&   0&   2097152&     100&0.94865200&  PASSED\\ 
diehard opso&   0&   2097152&     100&0.41217400&  PASSED\\ 
diehard oqso&   0&   2097152&     100&0.49075022&  PASSED \\
diehard dna&   0&   2097152&     100&0.78245172&  PASSED \\
diehard count 1sstr&   0&    256000&   100&0.59545874& PASSED\\ 
diehard count 1sbyt&   0&    256000& 100&0.44152512&  PASSED \\
diehard parking lot&   0&     12000&  100&0.99601277&   \rtext{WEAK}  \\
diehard 2dsphere&   2&      8000&     100&0.93564067&  PASSED\\ 
diehard 3dsphere&   3&      4000&     100&0.94103902&  PASSED \\
diehard squeeze&   0&    100000&     100&0.58677274&  PASSED\\ 
diehard sums&   0&       100&     100&0.76102130&  PASSED \\
diehard runs&   0&    100000&     100&0.11968119&  PASSED \\
diehard runs&   0&    100000&     100&0.51728489&  PASSED \\
diehard craps&   0&    200000&     100&0.86247445&  PASSED \\
diehard craps&   0&    200000&     100&0.97041678&  PASSED \\
marsaglia tsang gcd&   0&  10000000&  100&0.68088920&  PASSED \\
marsaglia tsang gcd&   0&  10000000&  100&0.20851577&  PASSED \\
sts monobit&   1&    100000&     100&0.15319982&  PASSED \\
sts runs&   2&    100000&     100&0.80047009&  PASSED  \\
sts serial&   1\ldots 16&    100000&     100&  0.11969309 --  0.98698338 &  PASSED \\
rgb bitdist&   1\ldots 12&    100000&     100& 0.07588499 -- 0.96001991 &  PASSED \\
rgb minimum distance&  2&  10000&    1000&0.19433266&  PASSED \\
rgb minimum distance&  3&  10000&    1000&0.45871020&  PASSED \\
rgb minimum distance&  4&  10000&    1000&0.53968425&  PASSED \\
rgb minimum distance&  5&  10000&    1000&0.82922021&  PASSED \\
rgb permutations&  2&    100000&     100&0.59213936&  PASSED \\
rgb permutations&   3&    100000&     100&0.77283570&  PASSED \\
rgb permutations&   4&    100000&     100&0.98499840&  PASSED \\
rgb permutations&   5&    100000&     100&0.47842741&  PASSED \\
rgb lagged sum&   0\ldots 32&   1000000&     100& 0.11439306 -- 0.99357917 &  PASSED \\
rgb kstest test&   0&     10000&    1000&0.81377150&  PASSED \\
dab bytedistrib&   0&  51200000&       1&0.45890415&  PASSED \\
dab dct& 256&     50000&       1&0.80469155&  PASSED \\
dab filltree&  32&  15000000&       1&0.22779072&  PASSED \\
dab filltree&  32&  15000000&       1&0.61178536&  PASSED \\
dab filltree2&   0&   5000000&       1&0.18837842&  PASSED \\
dab filltree2&   1&   5000000&       1&0.49022984&  PASSED \\
dab monobit2&  12&  65000000&       1&0.43688754&  PASSED \\
\hline
\toprule
\end{tabular*}
\end{table*}

\subsubsection{NIST SP800-22}
{As per NIST recommendations \cite{RukhinNIST2010tech},} we use $m=1500$ sequences of $1$ Mb each to assess the random numbers generated by the device. The tested sequences pass both the proportion and uniformity of the $p$-values assessments.  See Fig. \ref{Fig:NIST}.

 \begin{figure*}[!]
\centering
\includegraphics[width=0.45 \linewidth]{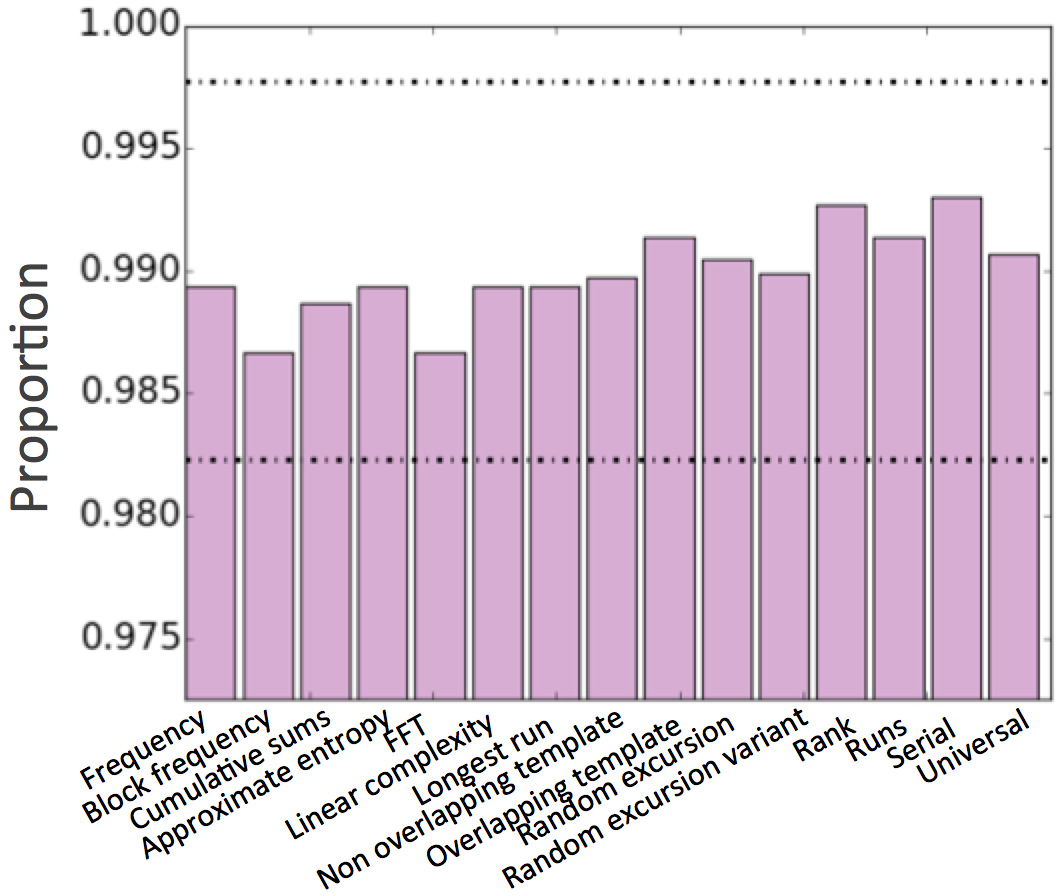}
\includegraphics[width=0.45 \linewidth]{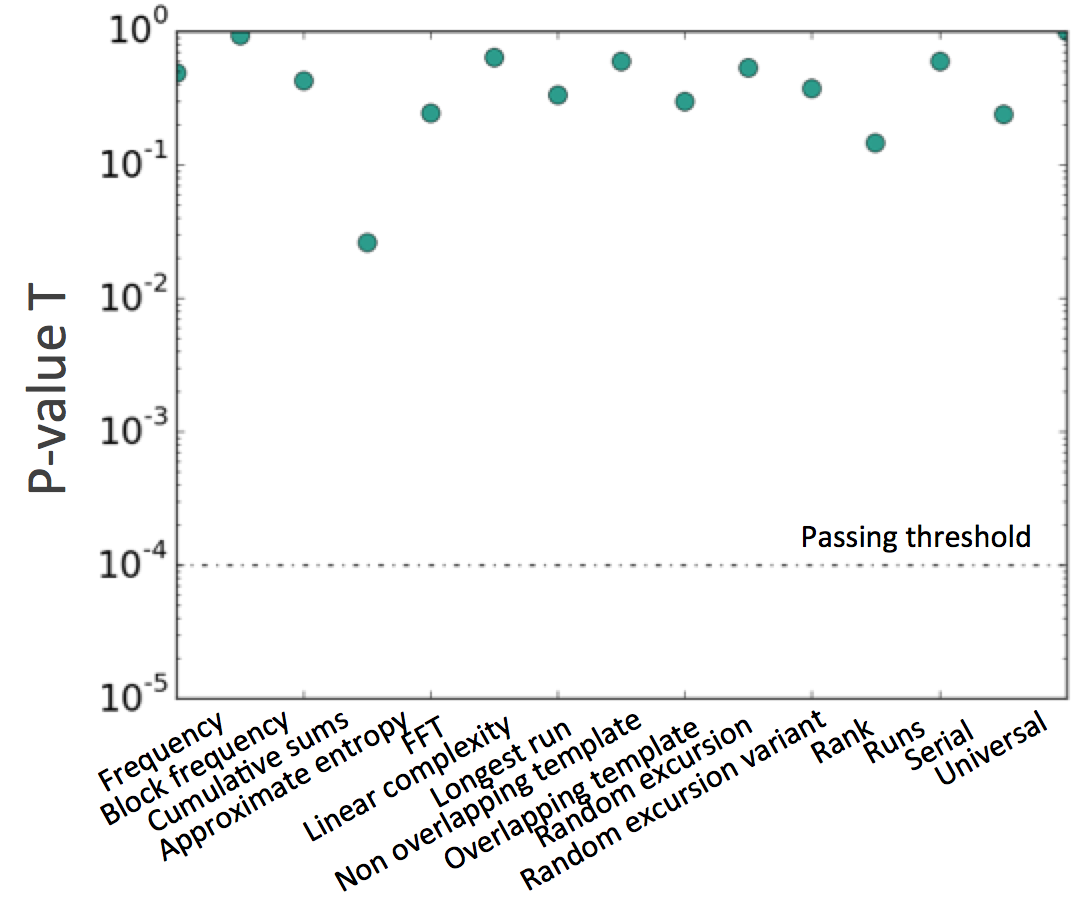}
\caption{ Summary of results for the NIST SP800-22 battery. The top figure depicts the results for the proportion test, i.e. evaluating how many times each test passes or fails (note that even a perfect random number generated is supposed to fail some tests from time to time). {Dot-dash lines indicate limits of acceptable behavior as recommended by NIST SP800-22.} The bottom picture represents the uniformity of the $p$-values assessment, which mainly quantifies how the outcomes of each statistic ($p$-values) are distributed along the (0,1) range. Both quantities satisfy the statistics for a perfect random number generator.  
\label{Fig:NIST} }
\end{figure*}

\subsubsection{Test U01 Alphabit battery}
We use the Alphabit battery following two testing strategies: (i) test many different sequences of a relatively small size, e.g. 300 files of 1 Gb each, and (ii) testing very long sequences, e.g. one file containing 1 Tb. Using (i), we can quantify how often the generator fails the Alphabit battery.  This is important because an ideal random number generator should fail with around $2\%$ probability. With (ii) we can test for weaker correlations/anomalies, below the statistical uncertainty of strategy (i). 

Results for strategy (i) are depicted in Fig. (\ref{Fig:testing}). We tested 1 Mb for k=1, 120 Mb for k=2, 500 Mb for k=3, and 1 Gb for k=4. In each case we test 300 sequences.  For strategy (ii) we tested a single $1$ Tb file, two $500$ Gb files, one $80$ Gb file, and two $64$ Gbfile for $k=3$ and {all tests were passed}. We followed the same criterion for evaluating the results as in \citeSOM{JacobsenMaster2014} in which regularities in commercial RNG systems were found for 64 Gb and above. 

\newcommand{\devP}{\Delta_{(p)}}
 \begin{figure*}[!h]
\centering
\includegraphics[width=0.45 \linewidth]{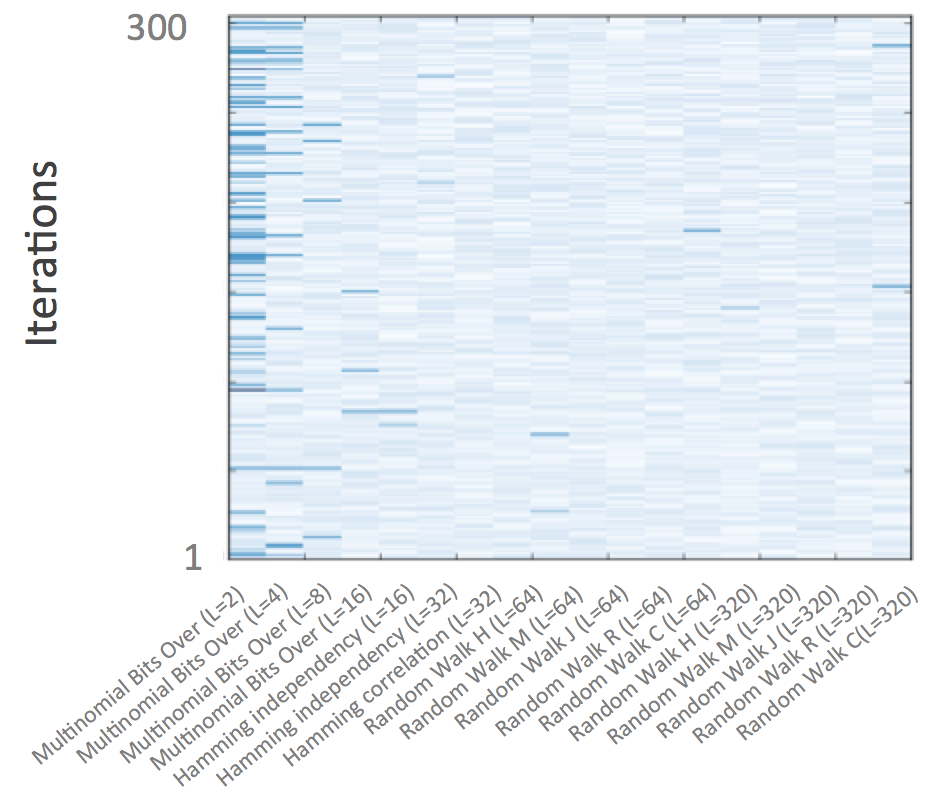}
\includegraphics[width=0.45 \linewidth]{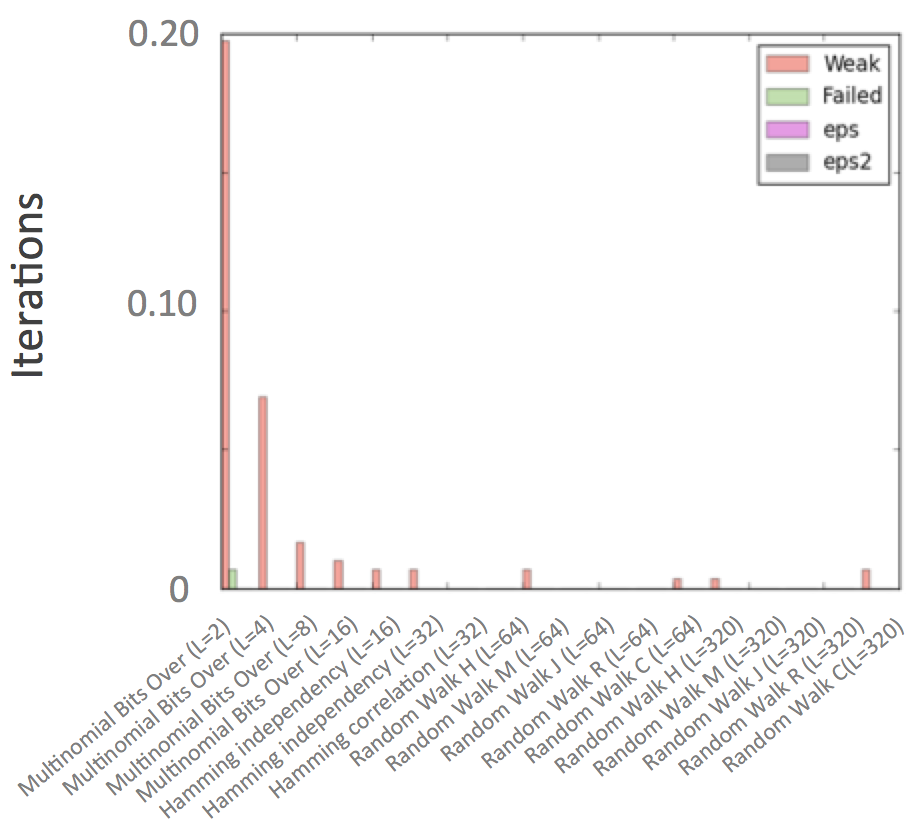}\\
\caption{\label{Fig:testing} {Statistical tests for $k=1$}. (Left)  represents the results for each of the 17 tests (horizontal axis) in the Alphabit battery for 300 iterations (vertical axis). A dark blue square represents a weak value is observed for that particular iteration and test. For each test, a $p$-value $p$ is computed and its deviation from zero or one $\devP = \min\{ 1-p , p \}$ is measured. If $\devP \geq 10^{-2}$ the test is considered to pass that statistic. In contrast, a test is considered inconclusive or weak when its deviation is $10^{-2}>\devP\geq 10^{-6}$, failed when $10^{-6} > \devP \geq 10^{-15}$,  ``eps,'' which implies catastrophic failure, when $10^{-15} > \devP \geq 10^{-300}$, and ``eps2'' when $\devP < 10^{-300}$. (Right) shows the frequency of obtaining weak, failed, eps, and eps2 p-values for each of the 17 tests of the alphabit battery.  
As shown, for k=1 the alphabit battery is able to find statistically significant anomalies, indicating the need for randomness extraction.
%As shown, no test is failed, neither with a small p-value nor with eps / eps2. For k=1, there are too many weak results, so it means the raw output is a weak random source. For k=2, the results are statistically acceptable, but it appears borderline. For k=3,4 (the k=4 case is not shown), the failure rate is {consistent with an ideal random source}.
}
\end{figure*}
 \begin{figure*}
\centering
\includegraphics[width=0.45 \linewidth]{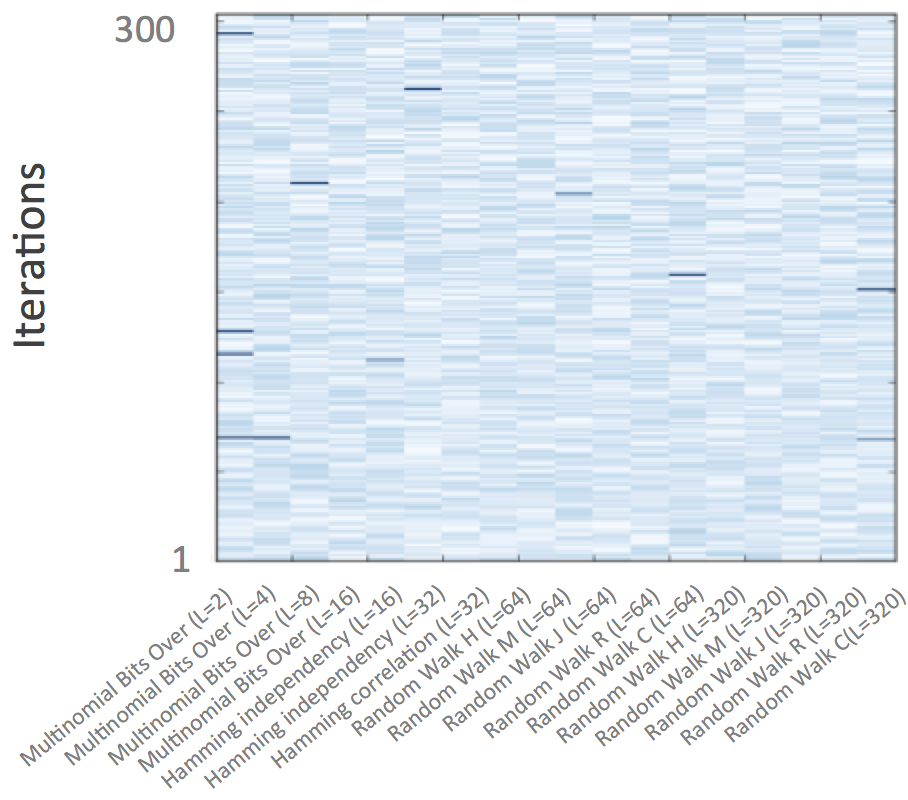}
\includegraphics[width=0.45 \linewidth]{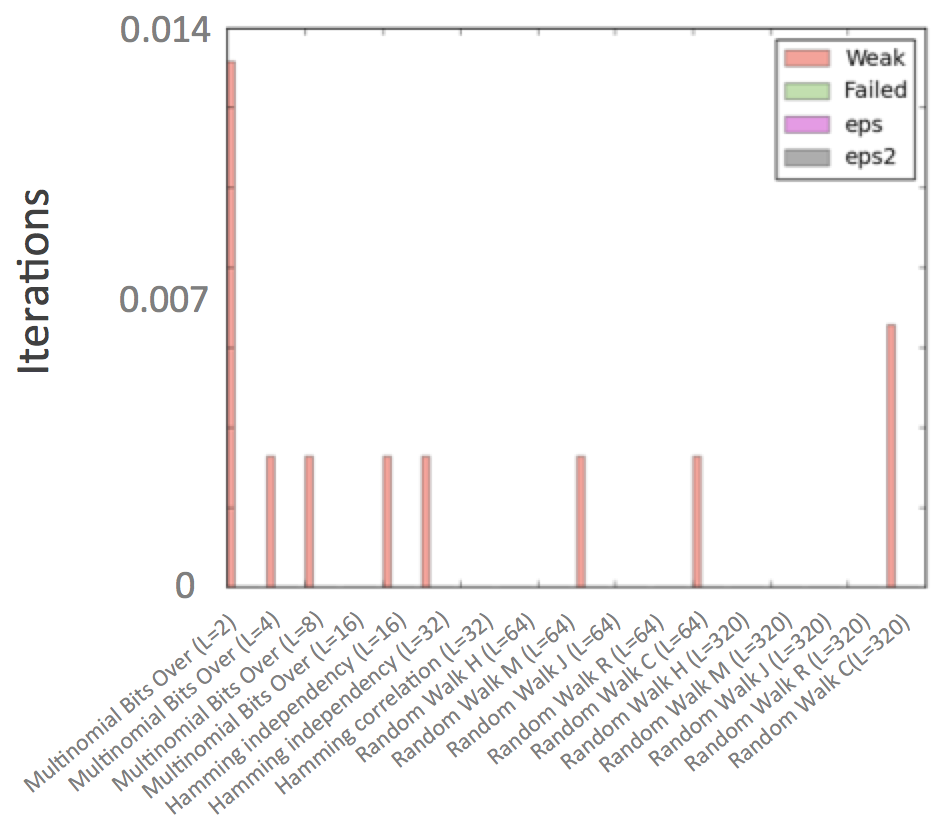}\\
\includegraphics[width=0.46 \linewidth]{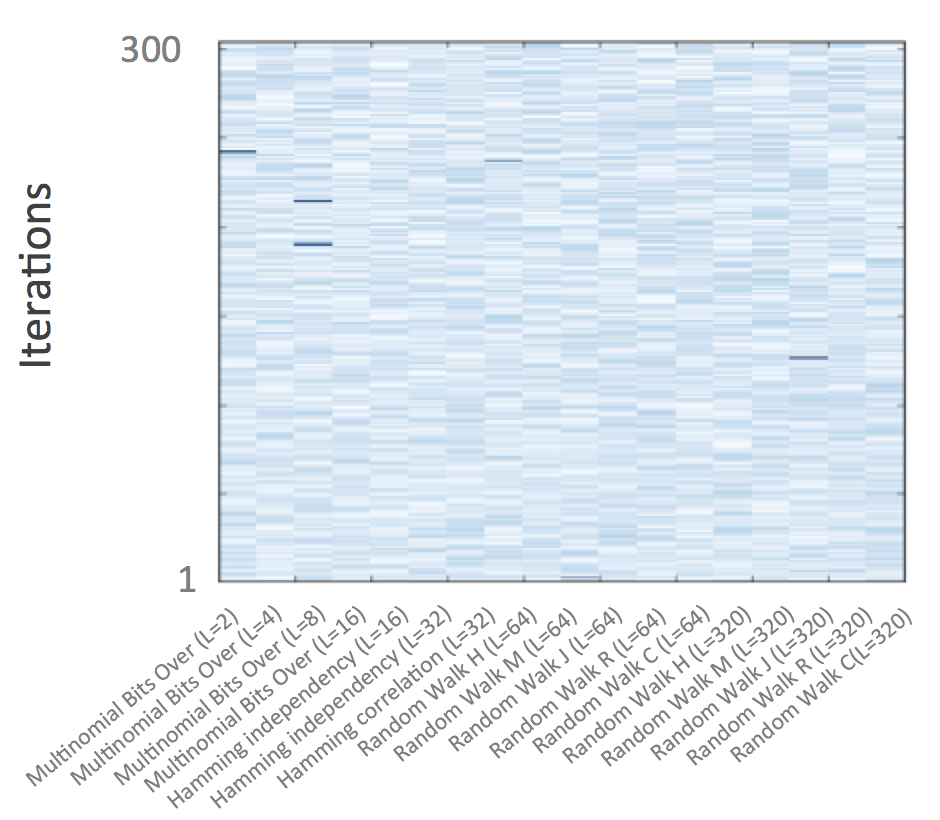}
\includegraphics[width=0.45 \linewidth]{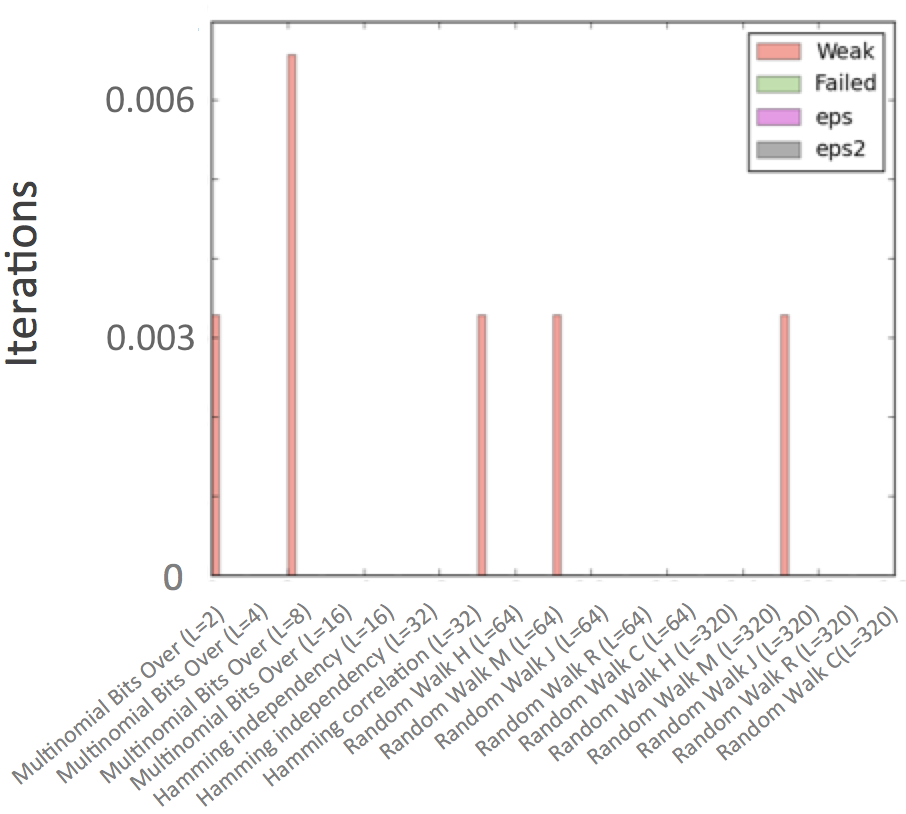}\\
\caption{\label{Fig:testing} {(Rows $1$ and $2$ correspond to the statistical tests for $k=2$ and $k=3$}, respectively. (Left column pictures) represent the results for each of the 17 tests (horizontal axis) in the Alphabit battery for 300 iterations (vertical axis). A dark blue square represents a weak value is observed for that particular iteration and test. For each test, a $p$-value $p$ is computed and its deviation from zero or one $\devP = \min\{ 1-p , p \}$ is measured. If $\devP \geq 10^{-2}$ the test is considered to pass that statistic. In contrast, a test is considered inconclusive or weak when its deviation is $10^{-2}>\devP\geq 10^{-6}$, failed when $10^{-6} > \devP \geq 10^{-15}$,  ``eps,'' which implies catastrophic failure, when $10^{-15} > \devP \geq 10^{-300}$, and ``eps2'' when $\devP < 10^{-300}$. (Right column pictures) show the frequency of obtaining weak, failed, eps, and eps2 p-values for each of the 17 tests of the alphabit battery.  For k=2, the results are statistically acceptable, but appear borderline. For k=3,4 (the k=4 case is not shown), the failure rate is consistent with an ideal randomness source.
%As shown, no test is failed, neither with a small p-value nor with eps / eps2. For k=1, there are too many weak results, so it means the raw output is a weak random source. For k=2, the results are statistically acceptable, but it appears borderline. For k=3,4 (the k=4 case is not shown), the failure rate is {consistent with an ideal random source}.
}
\end{figure*}

% The Kludge here is to

%\input{./SI.bbl}

%\begin{enumerate}
%%\begin{thebibliography}{1}
%
%\item{JacobsenMaster2014}
%Jakobsson, K.~S.
%\newblock Theory, methods and tools for statistical testing of pseudo and
%  quantum random number generators.
%\newblock Master's thesis, Link{\"o}pings {U}niversitet,  (2014).
%
%\end{enumerate}
%%\end{thebibliography}

%\input{./FQRNG151015SI.bbl}
%\input{./SI.bbl}

%\clearpage

\bibliographystyleSOM{apsrev4-1no-url}
\bibliographySOM{./freshqrng}

%\newsavebox\mytempbib
%\savebox\mytempbib{\parbox{\textwidth}{\bibliography{./freshqrng}}}

\end{document}